\documentclass[prb,aps,showpacs,twocolumn]{revtex4}

\usepackage{graphicx,color}
\usepackage{amsmath}
\usepackage{amssymb}
\usepackage{mathtools}
\usepackage{bm}
\usepackage{latexsym}
\usepackage{float}

\newcommand{\be}{\begin{equation}}
\newcommand{\ee}{\end{equation}} 
\newcommand{\lb}{\label}
\newcommand{\OL}{\overline}

\newcommand{\const}{({\rm const.})}

\newcommand{\ba}{{\bf a}}

\newcommand{\bk}{{\bf k}}

\newcommand{\br}{{\bf r}}
\newcommand{\bu}{{\bf u}}

\newcommand{\bx}{{\bf x}}

\newcommand{\bJ}{{\bf J}}

\newcommand{\bs}{{\bf s}}

\newcommand{\bomega}{{\mbox{\boldmath $\omega$}}}

\newcommand{\grad}{{\mbox{\boldmath $\nabla$}}}
\newcommand{\bdot}{{\mbox{\boldmath $\cdot$}}}

\newcommand{\bzed}{{\mbox{\boldmath $0$}}}

\DeclareMathOperator{\sinc}{sinc}
\definecolor{green}{rgb}{0,0.6,0}

\begin{document}

\title{The direct enstrophy cascade of two-dimensional soap film flows}

\author{M.K.~Rivera} \affiliation{Condensed Matter \& Thermal Physics
  Group,Los Alamos National Laboratory, Los Alamos, NM 87545}
\affiliation{Center for Nonlinear Studies, Los Alamos National Laboratory, Los
  Alamos, NM 87545}
\author{H.~Aluie} \affiliation{Center for Nonlinear Studies,
 Los Alamos National Laboratory, Los Alamos, NM 87545}
\author{R.E.~Ecke}  \affiliation{Center for
  Nonlinear Studies, Los Alamos National Laboratory, Los Alamos, NM 87545}

\date{\today}

\begin{abstract}
We investigate the direct enstrophy cascade of two-dimensional decaying turbulence in
a flowing soap film channel.
We use a coarse-graining approach that allows us to resolve the nonlinear dynamics and scale-coupling simultaneously in space and in scale. 
From our data, we verify an exact relation due to Eyink (1995) between traditional $3$rd-order structure function and the enstrophy flux obtained by coarse-graining. We also present experimental evidence that enstrophy cascades to smaller (larger) scales with a 60\% (40\%) probability, in support of theoretical predictions by Merilees \& Warn (1975) which  
appear to be valid in our flow owing to the ergodic nature of turbulence. We conjecture that their kinematic arguments break down in quasi-laminar 2D flows. We find some support for these ideas by
using an Eulerian coherent structure identification technique, which allows us to determine the effect of flow topology on the enstrophy cascade. A key finding is that ``centers'' are inefficient at transferring enstrophy between scales, in contrast to ``saddle'' regions which transfer enstrophy to small scales with high efficiency.
\end{abstract}

\maketitle
\section{Introduction \label{sec: Introduction}}
Two-dimensional turbulence has been studied extensively from theoretical (e.g., Ref. \cite{{Fjortoft53,Kraichnan67,Leith68,Batchelor69}}) and numerical (e.g., Ref.\cite{Chenetal03,Boffetta07,Bernardetal06}) standpoints, but flows in nature and in the laboratory are never exactly two-dimensional because there is always some degree of three-dimensionality. Many of the defining features of 2D turbulence, however, appear to be manifested in physical systems such as in oceanic, atmospheric, and planetary flows. This makes laboratory experiments of quasi-2D turbulence especially important to test agreement between physically realizable flows with idealized theory and numerics. 

Unlike in 3-dimensional flows, turbulence in 2-dimensions 
lacks the mechanism of vortex stretching, which implies that both energy 
and enstrophy (mean-square vorticity) are conserved. These two invariants
give rise to two separate cascades; an inverse cascade of energy to larger 
scales and a forward cascade of enstrophy to smaller scales. This explains why 
2D turbulent flows have a tendency to produce long-lived coherent structures at 
large-scales where energy accumulates.

Here we consider the direct enstrophy cascade for decaying grid turbulence\cite{BatchelorBook82}
where there are clear theoretical predictions, including an energy spectrum
$E(k) \sim k^{-3}$ (with logarithmic corrections), a forward constant flux of enstrophy $Z(k) = \eta$ where $\eta$ is a positive constant independent of wavenumber $k$ in the inertial scale-range, and precise
predictions for velocity and vorticity structure functions. Naturally occurring or experimentally realizable flows, however, inevitably deviate from idealized theoretical and numerical models upon which such predictions rest. For example, it has been observed that the presence of frictional linear drag in experiments of 2D turbulence perturbs the idealized 2D direct cascade picture by steepening the energy (and enstrophy) spectrum\cite{Tsangetal05,Boffetta07}, and by eliminating logarithmic signatures in spectra and structure functions \cite{Boffetta07,BoffettaEcke12}. 
Moreover, the theoretical predictions hold in the asymptotic limit of vanishing viscosity whereas realistic flows such as in our experiment are always at a finite Reynolds number.
Testing the extent to which theoretical predictions are  manifested in laboratory and naturally accessible settings is, therefore, essential in assessing their physical applicability and relevance.

Flowing soap films provide a very good experimental realization of the 
direct enstrophy cascade in 2D turbulence \cite{Kellayetal95,MartinPRL98,Rutgersetal01}.
Motivated by early seminal experiments \cite{Couderetal89,Gharibetal89}, a robust flowing soap
film apparatus was developed \cite{Kellayetal95,MartinPRL98} using single-point velocity 
measurements to characterize the turbulent state.  The introduction of 
particle-tracking velocimetry (PTV) to soap films allowed the measurement of the velocity {\it field} and
calculation of the corresponding vorticity \cite{Riveraetal98,Vorobieffetal99}.
Access to high-resolution velocity fields has enabled new analysis methods and diagnostics
to be used in investigating these complex flows.

One example is probing the role of vorticity in 2D turbulence. The physical mechanism responsible for the forward enstrophy cascade is vortex gradient stretching arising from mutual-interaction among vortices\cite{BoffettaEcke12}. Therefore, a thorough characterization of the enstrophy cascade must involve the diagnosis of vortices using, for example, coherent structure identification methods \cite{Okubo70,Weiss91,BasdevantPhilipovitch94,HuaKlein98} to investigate the topological and dynamical effects vorticity has on the flow and the cascade, as we show below.

Another example of PTV-enabled advanced diagnostics is the direct pointwise measurement of a cascade. A cornerstone idea of turbulence is the exchange of inviscid invariants, such as energy, between spatial scales \cite{Frisch_Turbulence} ---namely the nonlinear cascade process. This potent idea has had immense practical applications in turbulence closure, modeling, and prediction (e.g., Large Eddy Simulations\cite{MeneveauKatz00}). In this work, we analyze the cascade of enstrophy using a relatively novel method based on coarse-graining (or filtering) to measure the coupling between scales. The method (sometimes referred to as ``filter-space technique'') is rooted in the mathematical analysis of partial differential equations (called mollification, e.g., see Ref.\cite{Evans10}), and in LES \cite{Leonard74,Germano92}. The method was further developed mathematically by Eyink\cite{Eyink95a,Eyink95b,Eyink05} to analyze the physics of scale coupling in turbulence and has been refined and utilized in several fluid dynamics applications 
\cite{Piomellietal91,Liuetal94,Tao_2002_JFM,Riveraetal03,Chenetal03,AluieEyink09,AluieEyink10,AluieKurien11,KelleyOuellette11,AluieLiLi12}.

This paper is organized as follows. In Section \ref{sec:Experimental}, we describe the experimental apparatus and present standard metrics that characterize the turbulent state in our flow. Section \ref{sec:coarsegraining} provides a brief summary of the coarse-graining approach. We then show how the technique can be used to directly measure the coupling between scales at every flow location in Section \ref{sec:FilterEquations}. In Section \ref{sec:Results}, we apply the coarse-graining method to our experimental data and measure average fluxes, spatial distributions of those fluxes, and spatial correlations between coherent structures and the cascade. 
We conclude with Sec.\ref{sec:Conclusions} and defer some of details about the coarse-graining technique to an Appendix.

\section{Experimental Apparatus and Turbulent Quantities \label{sec:Experimental}}
Experimental measurements were carried out in a flowing soap-film channel, a
quasi-2D system in which decaying turbulence of low to moderate Reynolds
number can be generated ($10^2 \leq Re \leq 10^4$). This system was described in detail elsewhere\cite{Riveraetal98,Vorobieffetal99}. 

\begin{figure}[h]
  \includegraphics[width=3in]{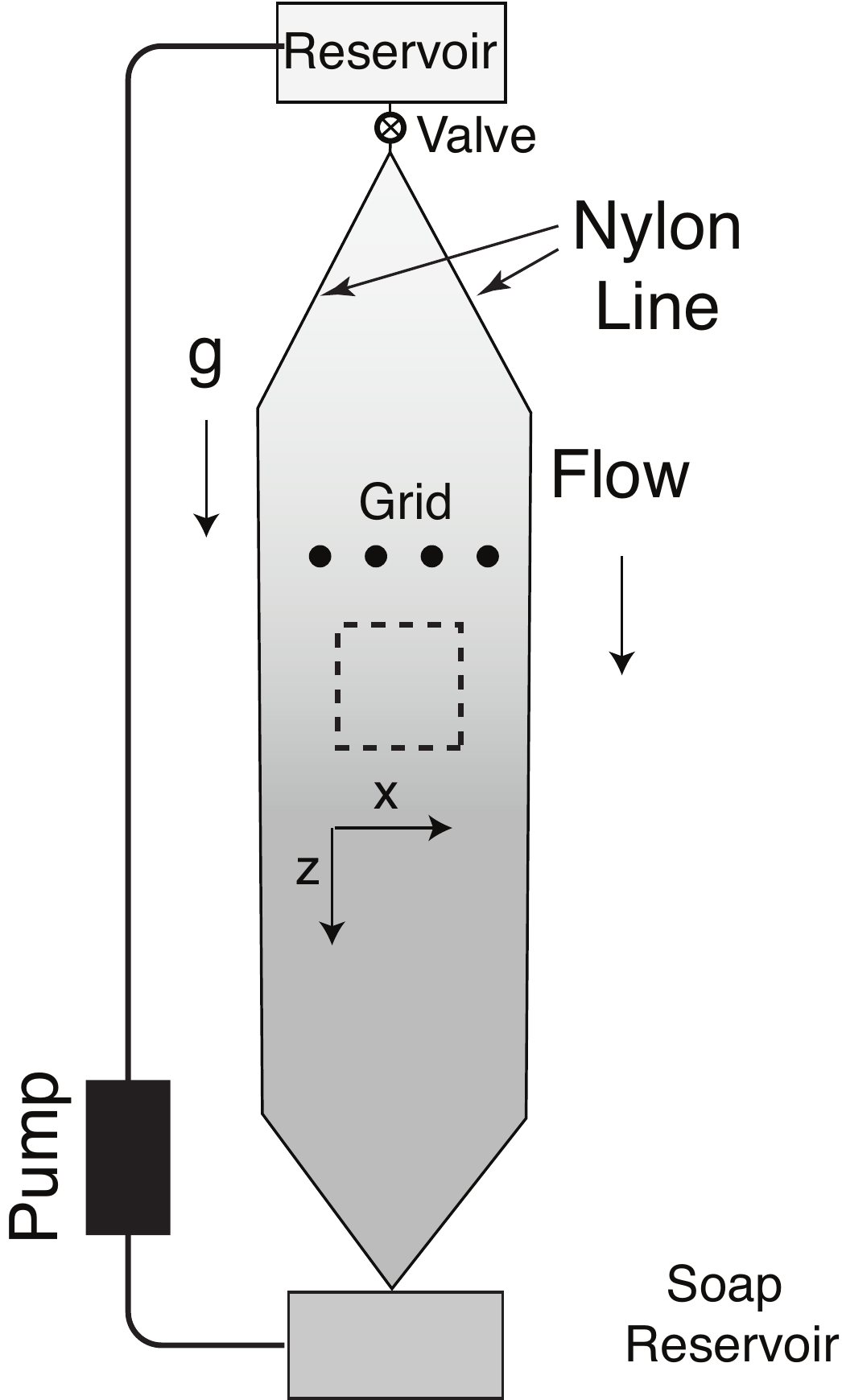}
  \caption{Schematic of a vertical soap-film channel. The film is constantly replenished using a pump, and the flow rate is adjusted with a valve from the top reservoir. The frame of the channel is typically made of nylon wires. The width of the channel can be comfortably varied in the range $1-10$ cm with a total height of $100-200$ cm.}
   \label{fig:experiment} 
\end{figure}

The surfactant-water solution, typically $2\%$ of commercial detergent in water, is continuously recirculated to the top of the channel by a pump. The flowing soap-film is suspended between the two nylon wires $5$ cm apart. The mean velocity depends on the volume flow rate and the tilt angle, $\theta$, of the channel. By varying $\theta$, the mean velocity can range from  0.5 m s$^{-1}$ to 4 m s$^{-1}$ and the soap film thickness can range between 1 and 30 $\mu$m. The resultant soap film can last for several hours.
All results reported here are from a channel inclined at an angle of $\theta = 75^{\circ}$ with respect to the vertical with a mean velocity of 
$U = 120$ cm/s and film thickness of about $10\ \mu$m.
 Turbulent flow is generated in the film channel by a 1D grid inserted in the film (see Figure \ref{fig:experiment}) with the separation between the teeth and their size determining the injection scale.

Using the empirical relationships
measured in \cite{VorobieffEcke99}, the films' kinematic viscosity was $\nu
\approx 0.03$ cm$^2$/s.  The turbulence generating grid consisted of rods of
$0.12$ cm diameter with $0.22$ cm spacing between the rods.  Thus, the
blocking fraction is around $0.3$, which is typical for turbulence in 2D soap
film flows
\cite{Gharibetal89,Kellayetal95,Riveraetal98,Rutgersetal01}.  The
resulting Reynolds number, $Re = UL/\nu$, was $880$ based on the mean-flow
velocity and an injection scale of $L_{\text{inj}} = 0.22$ cm.  
The Taylor-microscale Reynolds number was $R_\lambda\approx 200$ (for a root-mean-square velocity of about 25 cm/s) and a friction Reynolds number based on 
linear drag coefficient, $\alpha = 5$ sec$^{-1}$, due to friction with air, was $R_\alpha = U/L\alpha \approx 120$. 
Since we are primarily interested here, however, in the forward enstrophy cascade downscale of the forcing, 
$R_\alpha$ is not relevant for our study.
The turbulent
velocity, ${\mathbf u}({\mathbf x})$, and vorticity, $\omega({\mathbf x})$,
fields generated by the grid were obtained by tracking $3-5$ $\mu$m
polystyrene spheres (density approximately $1.05$ g/cc) within a $1.8 \times
1.8$ cm$^2$ region located $6$ cm downstream from the grid (20-30 eddy
rotation times)\cite{Ishikawa_2000_MST,Ohmi_2000_EiF}.  The particles were
illuminated with a double pulsed Nd:Yag laser and their images captured by a
12-bit, $2048\times2048$ pixel camera.  Around $3\times10^4$ particles were
individually tracked for each image pair and their velocities and local shears
were interpolated to a discrete $135 \times 135$ grid. One-thousand velocity
and vorticity fields were obtained in this way and were used to compute
ensemble averages of dynamical quantities described below. 
\begin{figure}
  \includegraphics[width=0.3\textwidth,height=0.2\textheight]{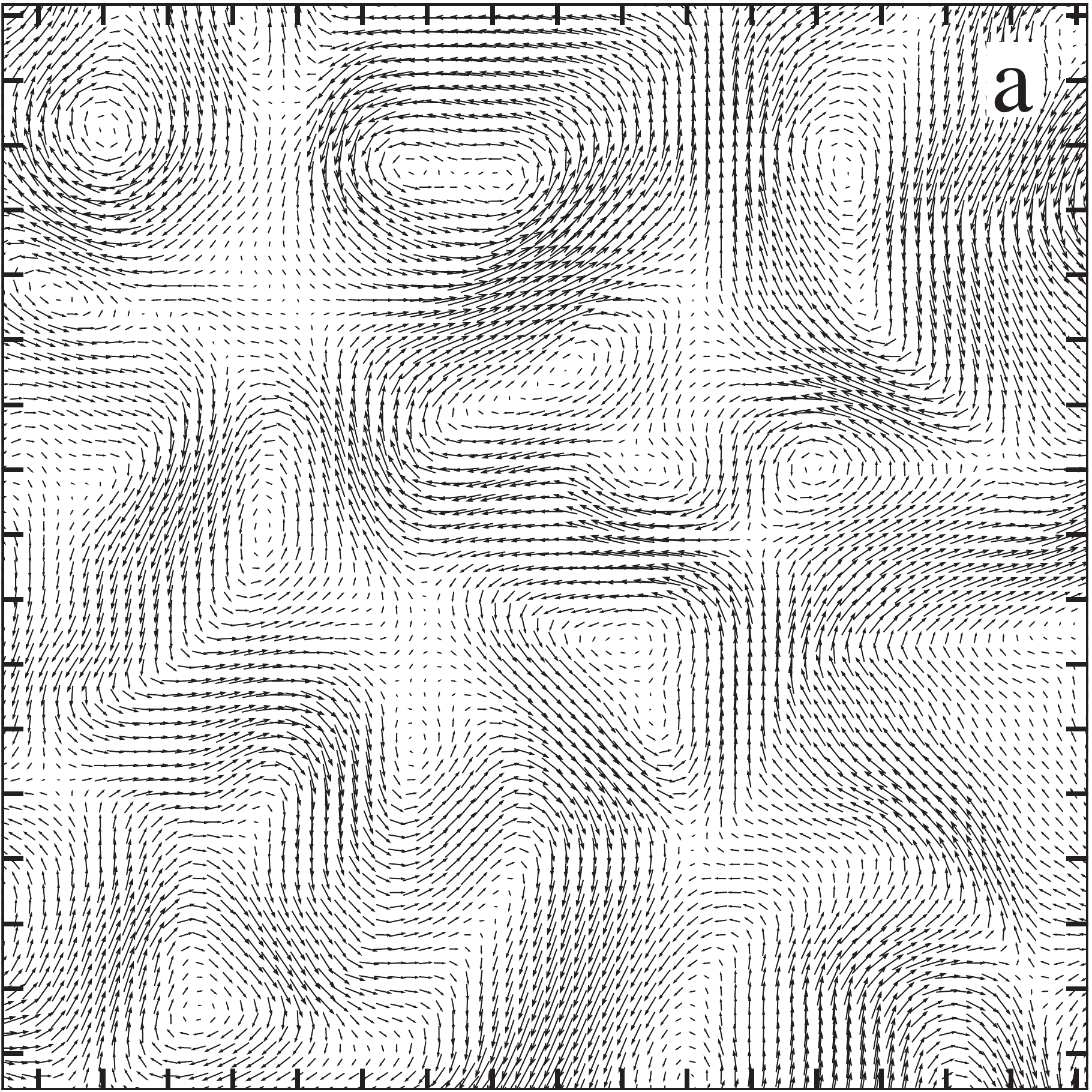}
  \includegraphics[width=0.3\textwidth,height=0.2\textheight]{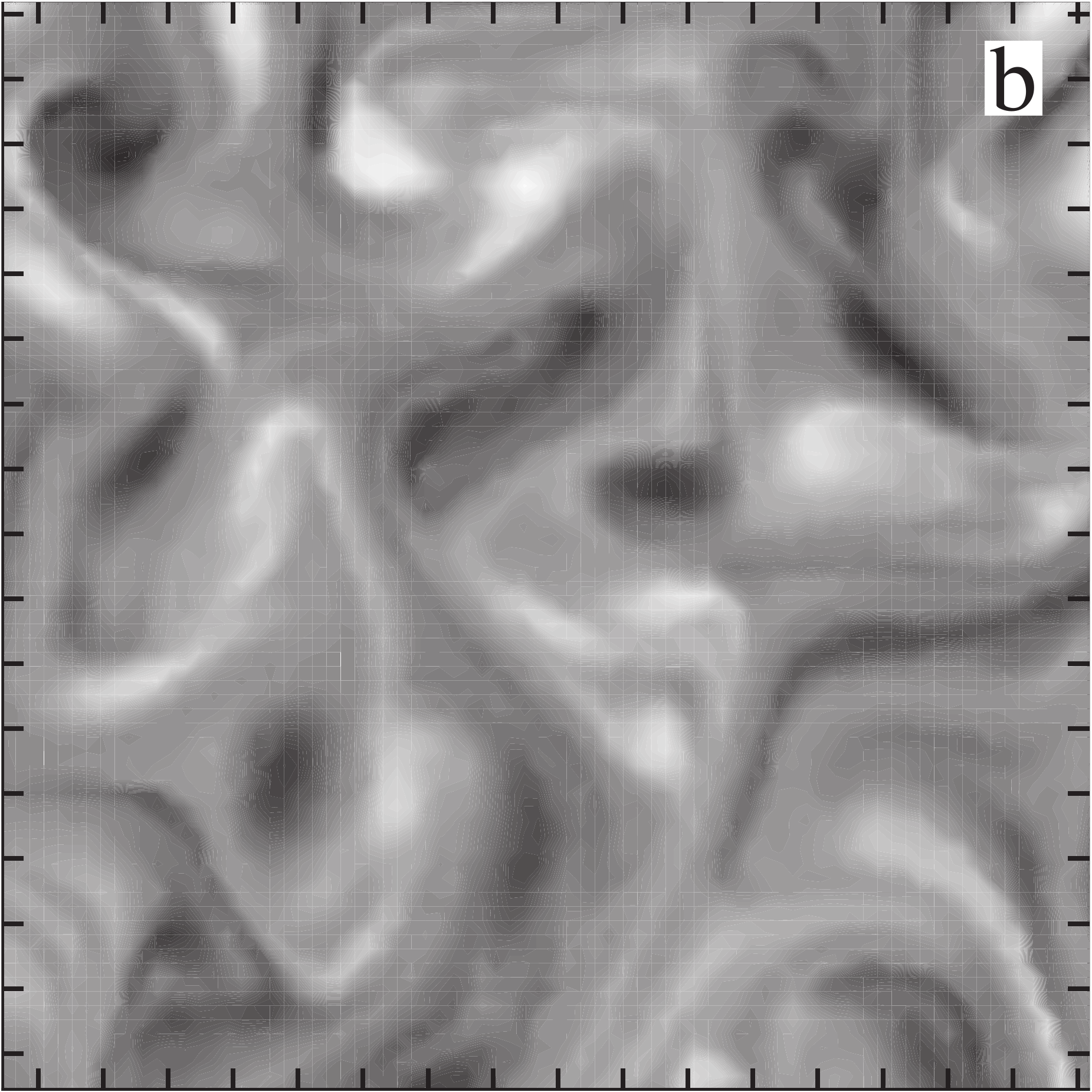}
  \caption{Typical a) velocity and corresponding b) vorticity fields
  obtained from the flowing soap film channel.  The hash marks represent
  $1$ mm increments. The mean flow is in the $-y$-direction and has been 
  subtracted out using Taylor's frozen-turbulence hypothesis. The top of the image is about 3 cm downstream from the
  energy injection grid.
   \label{fig:typical-fields} }
\end{figure}

\begin{figure}
 \includegraphics[width=3.2in]{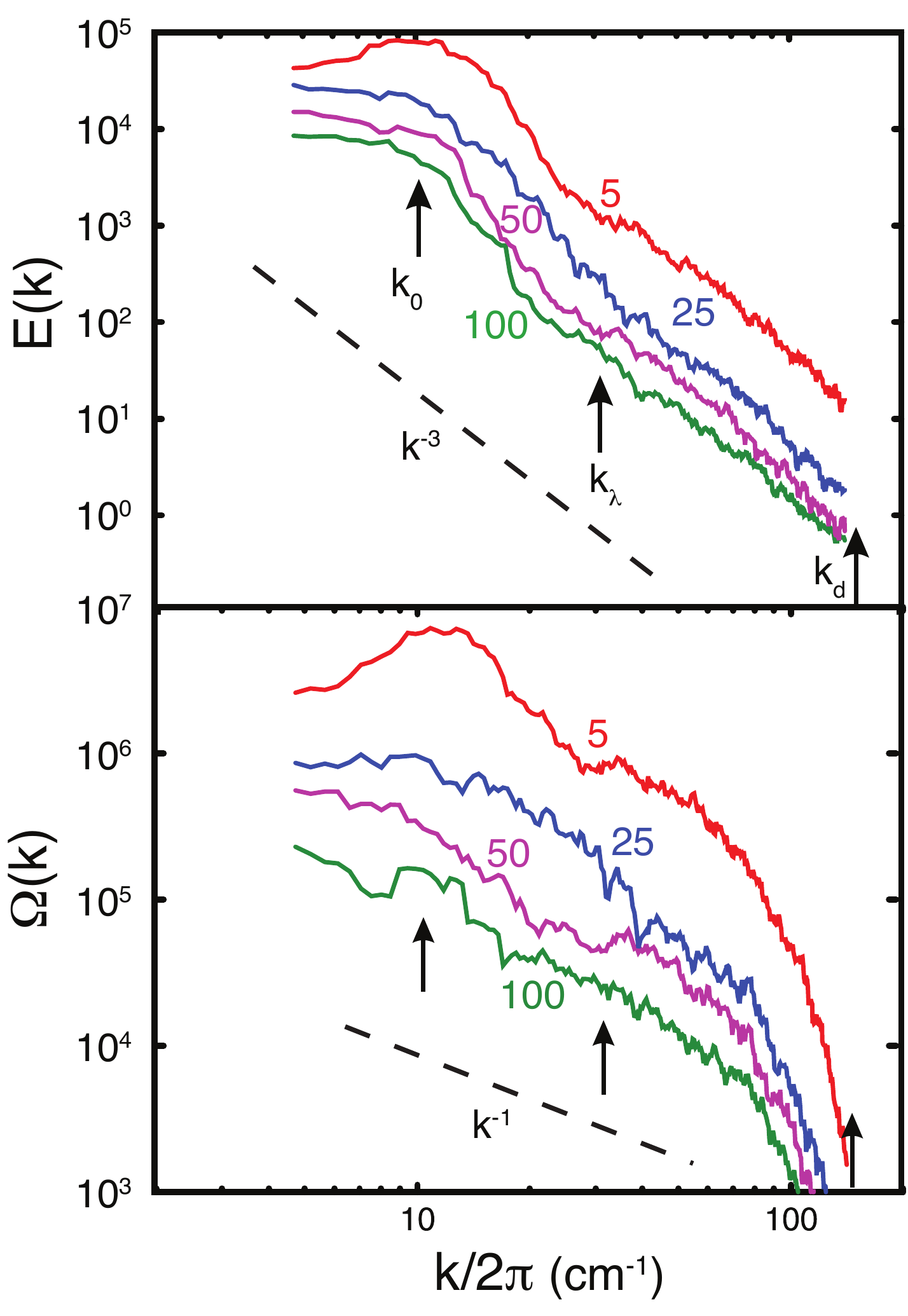}
   \caption{Spectra of energy, $E(k)$, (upper panel) and enstrophy, $\Omega(k)$, (lower panel) at various
  distances downstream from the grid: 5 cm (red), 25 cm (blue), 50 cm (violet), 100 cm (green).
  Arrows show the forcing wavenumber, $k_0$, Taylor micro-scale, $k_\lambda$, and viscous dissipation scale, $k_d$.
  Straight dashed lines are for reference and have slopes of $-3$ and $-1$ in the upper and lower panels, respectively. As turbulence develops downstream of the grid, spectra of energy and enstrophy exhibit putative power-law scaling 
  of $E(k)\sim k^{-3}$ and $\Omega(k)\sim k^{-1}$, expected in 2D turbulence; a 10\% steepening owing to friction is within the systematic uncertainty of the measurements. The exponential decay in $\Omega(k)$ 
  at $k>80$ is due to an effective filtering resultant from finite-differences.
  \label{fig:Spectra} }
\end{figure}

Typical velocity and vorticity fields are shown in Fig.~\ref{fig:typical-fields}. Measurements of velocity spectra in Figure \ref{fig:Spectra} show a power-law scaling over approximately one decade in wavenumber, $E(k)\sim k^{-\beta}$ with $\beta \approx 3$, as the flow decays away from the grid.
This is consistent with theoretical predictions \cite{Kraichnan67} and previous empirical observations\cite{Chenetal03,Boffetta07}.
Vorticity spectra in Figure \ref{fig:Spectra} also exhibit a power-law scaling consistent with $\Omega(k) \sim k^{-(\beta -2)} \approx k^{-1}$. Careful inspection of the data suggests that $\beta \gtrsim 3$, consistent with steepening resulting from frictional air drag \cite{Tsangetal05}.  Several experimental limitations are worth noting.  First, in the spectra there are a limited number of spatial points, of order 100, in each direction.  Thus, spectral slope differences of 10-20\% may arise from different choices of window functions.  Second, the exponential decay in $\Omega(k)$ at $k>80$ is a result of finite difference operation when computing vorticity, which effectively acts as filtering (e.g., see \cite{SagautBook00}). Taking into account these systematic uncertainties, we conclude that $\beta = 3 \pm 0.5$, consistent with theoretical predictions but also allowing the expected steepening for frictional drag.

\section{The Coarse-graining Approach \label{sec:coarsegraining}}

The key analysis method we use is a ``coarse-graining'' or ``filtering'' approach to analyzing scale interactions in complex flows.
The method is rooted in a standard technique in partial differential equations
and distribution theory (e.g., see Ref.\cite{Evans10}) but was introduced to the field of turbulence by
Leonard\cite{Leonard74} and Germano\cite{Germano92} in the context of Large Eddy Simulation modelling.
Eyink\cite{Eyink95a,Eyink05} developed the formalism mathematically to analyze the fundamental physics of scale coupling
in turbulence, which was later applied to numerical and experimental flows of 2D turbulence
\cite{Riveraetal03,Chenetal03,Chenetal06}. More recently, the approach was further refined and extended to magnetohydrodynamic \cite{AluieThesis,AluieEyink10}, geophysical \cite{AluieKurien11}, and compressible \cite{Aluie11,AluieLiLi12,Aluie13} flows. 

The method itself is simple. For any field $\ba(\bx)$, a ``coarse-grained'' or (low-pass) filtered field, which contains modes
at length-scales $>\ell$, is defined as
\be
\OL \ba_\ell(\bx) = \int d^{\scriptsize\mbox{n}}\br~ G_\ell(\br) \ba(\bx+\br),
\lb{filtering}\ee
where $n$ is the spatial dimension, $G(\br)$ is a normalized convolution kernel, $\int d^{\scriptsize\mbox{n}}\bs ~G(\bs)=1$, for dimensionless $\bs$. 
The kernel can be any real-valued function which decays sufficiently rapidly for large $r$. It is further assumed
that $G$  is centered, $\int d^{\scriptsize\mbox{n}} \bs ~\bs \,G(\bs) = \bzed$, and with the main support
in a ball of unit radius, $\int d^{\scriptsize\mbox{n}} \bs ~|\bs|^2 G(\bs) = \mathcal{O}(1)$. 
Its dilation in an $n$-dimensional domain, $G_\ell(\br)\equiv \ell^{-{\scriptsize\mbox{n}}} G(\br/\ell)$, will share these properties 
except that its main support will be in a ball of radius $\ell$. If $G(\bs)$ is also non-negative,
then operation (\ref{filtering}) may be interpreted as a local space average \footnote{Note that $G(\br)$ can be
chosen so that	both it and its Fourier transform, $\hat{G}(\bk)$, are positive and 
infinitely differentiable, with $\hat{G}(\bk)$ also compactly supported inside a 
ball of radius $1$ about the origin in Fourier space and with $G(\br)$ decaying 
faster than any power $r^{-p}$ as $r\to\infty$. See for instance Appendix A in 
Ref.\cite{EyinkAluie09} for explicit examples.}. An example of such a kernel in 1-dimension is the Gaussian 
function, $G(r) = \frac{1}{\sqrt{2\pi}}e^{-r^2/2}$. 

We can also define a complementary high-pass filter which retains only modes
at scales $<\ell$ by
\be  \ba^{'}_\ell(\bx) = \ba(\bx)-\OL\ba_\ell(\bx).
\lb{high-pass}\ee
In the rest of our paper, we drop subscript  $\ell$ whenever there is no risk of 
ambiguity.

From the dynamical equation of field $\ba(\bx)$,
coarse-grained equations can then be written to describe the evolution of 
$\OL{\ba}_\ell(\bx)$ at every point $\bx$ in space and at any instant of time. Furthermore,
the coarse-grained equations describe flow at scales $>\ell$, for arbitrary
$\ell$. The approach, therefore, allows for the simultaneous 
resolution of dynamics \emph{both in scale and in space} and admits
intuitive physical interpretation of various terms in the coarse-grained balance as we elaborate below

Moreover, coarse-grained equations describe the large-scales whose
dynamics is coupled to the small-scales through so-called \emph{subscale} 
or \emph{subgrid} terms (see, for example, Eq. (\ref{Turbulentstress})). These terms depend inherently on the unresolved 
dynamics which has been filtered out. 
The approach thus quantifies the coupling between different scales and may be used to extract
certain scale-invariant features in the dynamics. We utilize it here to investigate the transfer of enstrophy across scales in our 2D soap film flow experiments.

\section{Analyzing Nonlinear Scale Interactions in 2D flows \label{sec:FilterEquations}}

The simplest model to describe flow in soap films is that of 2D Navier-Stokes,
\be \partial_t \bu + (\bu\bdot\grad)\bu = -\grad p + \nu \nabla^2\bu -\alpha \,\bu,
\lb{VelocityEq}\ee
where $\bu$ is the velocity and is incompressible, $\grad\bdot\bu=0$, $p$ is pressure, $\nu$ is kinetic shear viscosity, and $\alpha$ is a linear drag coefficient owing to friction between the soap film and the air. An equation equivalent to (\ref{VelocityEq}) is that of vorticity, $\omega = \partial_x u_y - \partial_y u_x$:
\be \partial_t \omega + (\bu\bdot\grad)\omega =  \nu \nabla^2\omega -\alpha \,\omega,
\lb{VorticityEq}\ee
which does not contain the vortex stretching term, $(\bomega\bdot\grad)\bu$, that is critical in the dynamics of 3D flows.

Eq. (\ref{VorticityEq}) implies that inviscid and unforced 2D flows (with $\nu=\alpha=0$) are constrained by the conservation of vorticity following material flow particles,
\be D_t \omega = \partial_t \omega + (\bu\bdot\grad) \omega = 0.\ee

In addition to the Lagrangian conservation of vorticity, the flow is constrained by the global conservation of 
energy, $E = \langle|\bu|^2\rangle/2$, and enstrophy, $\Omega = \langle \omega^2 \rangle/2$, such that
\be \frac{d}{dt} E= \frac{d}{dt} \Omega = 0.
\ee
It is worth noting that 2D flows have an infinite set of Lagrangian invariants, $\omega^n$, and global invariants, $\langle\omega^n\rangle$,
for any integer $n\ge1$. These are usually called ``Casimirs'' (see for e.g., Refs.\cite{KraichnanMontgomery80,BouchetVenaille12}).
Our analysis in this study, however, will be restricted to vorticity and enstrophy.

There is agreement among theory\cite{Fjortoft53,Kraichnan67,Leith68,Batchelor69}, numerics\cite{Chenetal03,Boffetta07}, and experiments\cite{Riveraetal03,BoffettaEcke12} that in 2D flows, there are two distinct scale-ranges. Over a range of scales larger than that of injection, called the inverse cascade range,  energy is transferred upscale on \emph{average}. Over another range of scales smaller than that of injection, called the forward cascade range, enstrophy is transferred downscale on \emph{average}.
Beyond mere averages, however, the coarse-graining approach can yield a wealth of spatial information and statistics about the nonlinear coupling involved in transferring energy and enstrophy within the flow. The main focus of our paper will thus be on the enstrophy cascade.

\subsection{Coarse-grained Equations}
The filtering operation (\ref{filtering}) is linear and commutes with space and time derivatives. 
Applying it to the equation (\ref{VelocityEq}) yields coarse-grained equation for $\OL{\bu}_\ell(\bx)$ that describes the flow at scales larger than $\ell$ at every point $\bx$ in space and at any instant of time:
\begin{eqnarray} 
\partial_t \OL\bu_\ell + (\OL\bu_\ell\bdot\grad)\OL\bu_\ell &=& -\grad \OL{p}_\ell -\grad\bdot\OL\tau_\ell(\bu,\bu)+ \nu \nabla^2\OL\bu_\ell -\alpha \,\OL\bu_\ell,\hspace{.5cm}\lb{largeVelocityEq}\\
\grad\bdot\OL\bu_\ell&=&0,\nonumber
\end{eqnarray}
where the \emph{subgrid stress}
\be\OL\tau_\ell(\bu,\bu)\equiv \OL{(\bu\bu)}_\ell - \OL{\bu}_\ell~ \OL{\bu}_\ell,\lb{Turbulentstress}\ee
is a ``generalized 2nd-order moment'' \cite{Germano92}. It is easy to see that filtered equation (\ref{largeVelocityEq}) is similar to the original Navier-Stokes equation (\ref{VelocityEq}) but with an addition of the subscale term accounting for the influence of eliminated fluctuations at scales $<\ell$.  Since we have knowledge of the dynamics at all relevant scales in the system, the subscale term can be calculated exactly at every point $\bx$ in the domain and at any instant in time t. 

Alternatively, we can apply the filtering operation to the vorticity equation (\ref{VorticityEq}) which yields a coarse-grained equation for $\OL{\omega}_\ell(\bx)$:
\be \partial_t \OL\omega_\ell + (\OL\bu_\ell\bdot\grad)\OL\omega_\ell =  -\grad\bdot\OL\tau_\ell(\bu,\omega) + \nu \nabla^2\OL\omega_\ell -\alpha \,\OL\omega_\ell,
\lb{largeVorticityEq}\ee
where
\be\OL\tau_\ell(\bu,\omega)\equiv \OL{(\bu \,\omega)}_\ell - \OL{\bu}_\ell~ \OL{\omega}_\ell.\lb{Vorticitystress}\ee

\subsection{Large-scale energy budget}
From eq. (\ref{largeVelocityEq}) it is straightforward to derive the kinetic energy budgets for the large-scales $(>\ell)$, which reads
\begin{eqnarray} 
\partial_t \frac{|\OL\bu_\ell|^2}{2} + \grad\bdot\bJ_\ell = -\Pi_\ell  -\nu|\grad\OL\bu_\ell |^2 - \alpha |\OL\bu_\ell|^2,
\lb{largeKE}
\end{eqnarray}
Terms $\nu|\grad\OL\bu|^2$ and $\alpha |\OL\bu_\ell|^2$ are viscous dissipation and linear dissipation from air drag, respectively, 
acting directly on scales $>\ell$. The remaining terms in eq. (\ref{largeKE}) are defined as
\begin{eqnarray}
&J_j(\bx)&  = ~~ \OL{u}_j \frac{|\OL\bu|^2}{2} + \OL{p}\, \OL{u}_j+ \OL{u}_i~\OL\tau(u_i,u_j) - \nu\partial_j\frac{{|\OL\bu|^2}}{2}\lb{KEtransport}\\[0.3cm]
&\Pi_\ell(\bx)& = ~  -\partial_j\OL{u}_i  ~\OL\tau(u_i,u_j) = -\OL{S}_{ij}  ~\OL\tau(u_i,u_j) ~~ \lb{KEflux}
\end{eqnarray}
Space transport of large-scale energy is $\bJ_\ell(\bx)$, which only acts to redistribute the energy in space due to large-scale flow (first term in eq. (\ref{KEtransport})), large-scale pressure (second term in eq. (\ref{KEtransport})), turbulence (third term in eq. (\ref{KEtransport})), and viscous diffusion (last term in eq. (\ref{KEtransport})).
As we interpret these transport terms, they are not involved in the transfer of energy \emph{across} scales. In a statistically homogenous flow, the spatial average of such transport vanishes, $\langle\grad\bdot\bJ_\ell(\bx)\rangle =0$. 

Term $\Pi_\ell(\bx)$ is subgrid scale (SGS) kinetic energy flux. It involves the action of the large-scale velocity gradient, $\grad\OL\bu(\bx)$, against subscale fluctuations. Since the subgrid stress is a symmetric tensor, $\OL{\tau}(u_i,u_j) = \OL{\tau}(u_j,u_i)$, the SGS flux $\Pi_\ell(\bx)$ can be rewritten in terms in the large-scale symmetric strain tensor, $$\OL{S}_{ij} = \frac{1}{2}(\partial_j \OL{u}_i + \partial_i \OL{u}_j).$$ 
$\Pi_\ell(\bx)$ acts as a sink in the large-scale budget (\ref{largeKE}) and
accounts for energy transferred from scales $>\ell$ to smaller scales $<\ell$ at any point $\bx$ in the flow at any given instant in time.
Note that $\Pi_\ell(\bx)$ is Galilean invariant such that the rate of energy cascading at any position $\bx$ does not depend on an observer's velocity. Galilean invariance of the SGS flux was emphasized in several studies\cite{Speziale85,Germano92,AluieKurien11}, and shown to be necessary for scale-locality of the cascade\cite{EyinkAluie09,AluieEyink09}. There are non-Galilean invariant terms in budget (\ref{largeKE}) but, as is physically expected, they are all associated with spatial transport, $\bJ_\ell(\bx)$, and linear drag.

\subsection{Large-scale enstrophy budget}
Similar to the large-scale energy budget, one can derive a large-scale enstrophy budget from eq. (\ref{largeVorticityEq}) for the large-scales $(>\ell)$, which reads
\begin{eqnarray} 
\partial_t \frac{|\OL\omega_\ell|^2}{2} + \grad\bdot\bJ^{\Omega}_\ell = -Z_\ell  -\nu|\grad\OL\omega_\ell |^2 - \alpha |\OL\omega_\ell|^2,
\lb{largeEnstrophy}
\end{eqnarray}
Terms $\nu|\grad\OL\omega |^2$ and $\alpha |\OL \omega_\ell|^2$ are dissipation of large-scale enstrophy due to viscosity and linear drag, respectively, acting directly on scales $>\ell$. We also have in eq. (\ref{largeEnstrophy}) terms:
\begin{eqnarray}
&J^{\Omega}_j(\bx)&  = ~~ \OL{u}_j \frac{|\OL\omega|^2}{2} + \OL{\omega}~\OL\tau(\omega,u_j) - \nu\partial_j\frac{{|\OL\omega|^2}}{2}\lb{Enstrophytransport}\\[0.3cm]
&Z_\ell(\bx)& = ~  -\partial_j\OL{\omega}  ~\OL\tau(\omega,u_j)  ~~ \lb{Enstrophyflux}
\end{eqnarray}
similar to the energy budget, $\bJ^{\Omega}_\ell(\bx)$, acts to redistribute large-scale enstrophy in space due to the large-scale flow (first term in eq. (\ref{Enstrophytransport})), turbulence (second term in eq. (\ref{Enstrophytransport})), and viscous diffusion (last term in eq. (\ref{Enstrophytransport})).

\begin{figure*}[h]
   \includegraphics[height=6in,width=6in]{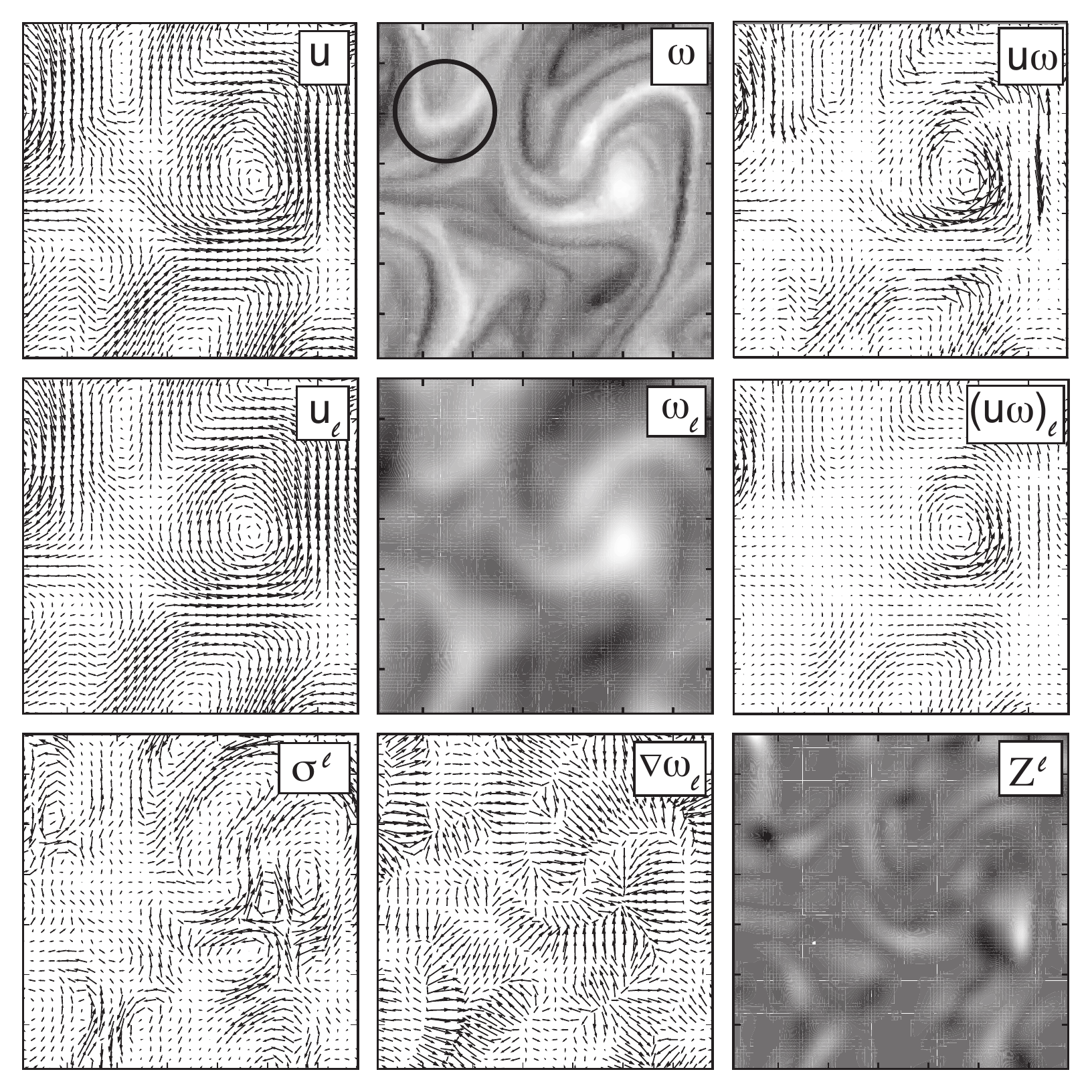}
   \caption{ 
   Obtaining the scale-to-scale enstrophy flux $Z_{\ell}(\bx)$ for a 
  velocity field obtained from the soap film over a $1 \text{cm} \times 1 \text{cm}$ area. The ticks at the edge of images above are 0.1 cm apart.  The filter function, $G_\ell(\br)$,
  used was Gaussian with $\ell=0.2$ cm. The averaging domain is shown above as a circle of diameter $0.2$ cm in
  the image of unfiltered $\omega$.  
  Row 1: Unfiltered velocity, ${\bf u}$, vorticity, $\omega$, and vorticity transport ${\bf u}\, \omega$. 
  Row 2: Filtered velocity, $\OL\bu_\ell$, vorticity, $\OL\omega_\ell$, and vorticity transport, $\OL{(\bu\,\omega)}_\ell$.  
  Row 3: The subgrid vorticity transport vector $\sigma^{(\ell)}=\OL\tau_\ell(\bu,\omega)$, large scale vorticity gradient 
  $\grad\OL\omega_\ell$ and the scale-to-scale enstrophy transfer $Z_{\ell}(\bx)$.   
   \label{fig:enstrophy-flux-pictorial} }
\end{figure*}

Term $Z_\ell(\bx)$ is subgrid scale (SGS) enstrophy flux which acts as a sink in the large-scale budget (\ref{largeEnstrophy}) and
accounts for the pointwise enstrophy transferred from scales $>\ell$ to smaller scales $<\ell$ (see Fig. \ref{fig:enstrophy-flux-pictorial}). Similar to $\Pi_\ell(\bx)$, the enstrophy flux, $Z_\ell(\bx)$, is Galilean invariant.

Figure \ref{fig:enstrophy-flux-pictorial} illustrates how the enstrophy flux, 
$Z_\ell(\bx)$, is calculated from our soap film experimental data. 
It is based on the following straightforward steps: 
(i) filter the velocity and vorticity fields, $\OL\bu_\ell(\bx)$ and $\OL\omega_\ell(\bx)$, 
(ii) filter the quadratic nonlinearity, $\OL{(\bu\,\omega)}_\ell$,
(iii) obtain the forces exerted on vorticity at scales larger than $\ell$ owing to nonlinear contributions from 
 fluctuations at scales smaller than $\ell$ by subtracting large-scale sweeping effects 
 from the quadratic nonlinearity,
$\OL\tau_\ell(\bx)(\bu,\omega)=\OL{(\bu\omega)}_\ell - \OL\bu_\ell\OL\omega_\ell$,
(iv) compute the gradient of large-scale vorticity, $\grad\OL\omega_\ell(\bx)$,
(v) the enstrophy flux results from the action of large-scale vorticity gradient against subscale fluctuations,
$Z_\ell(\bx) = -\grad\OL\omega_\ell\bdot\OL\tau_\ell(\bu,\omega)$.

For this calculation the filter function was a Gaussian with Fourier-space definition
\begin{equation}
\widehat{G}_{\ell}({\bf k})=e^{-\frac{|{\bf k}|^2}{k_\ell^2}},
\label{eq:GaussianKernel}
\end{equation}
where $k_\ell \equiv 2\pi/\ell$.
In Appendix A, we consider properties of different filtering kernels used and show that
our results are not sensitive to the kernel choice. We also discuss the criterion used here
when filtering in the presence of boundaries and discuss the effect of limited data resolution.

\clearpage
\section{Results\label{sec:Results}}

Our main results concern the transfer of enstrophy from large to small scales which is expected in the limit of very large Reynolds number to support a constant enstrophy flux.  For the modest $Re$ reported here, injection and dissipation are not negligible so the measured enstrophy flux is not precisely constant.  Nevertheless, the turbulence properties that we measure are consistent with the key features predicted by Kraichnan-Bachelor theory \cite{Fjortoft53,Kraichnan67,Leith68,Batchelor69} modified slightly by the addition of air drag friction.  In particular, for example, the enstrophy flux is positive, a necessary condition for a forward enstrophy cascade.  We present those results first. A secondary topic concerns the flux of energy which for forced, dissipative 2D turbulence is expected to form (in the large $Re$ limit) an inverse energy cascade (constant flux of energy to larger spatial scales).  For decaying 2D turbulence in our soap film apparatus, theory provides no definitive guidance.  Nevertheless, there is no reason to rule out an upscale \emph{transfer} of energy for the decaying turbulence scenario as discussed below.  As we show, energy accumulates in lower $k$ modes but there is no evident signature of inverse \emph{cascade} in the spectra for $E(k)$ or $\Omega(k)$.  We discuss below the implications for inverse energy transfer; first we concentrate on the forward enstrophy cascade.

\subsection{Space-averaged terms in the enstrophy budget\lb{sec:SpaceAveragedTerms}}

\begin{figure}[h]
  \includegraphics[width=3in]{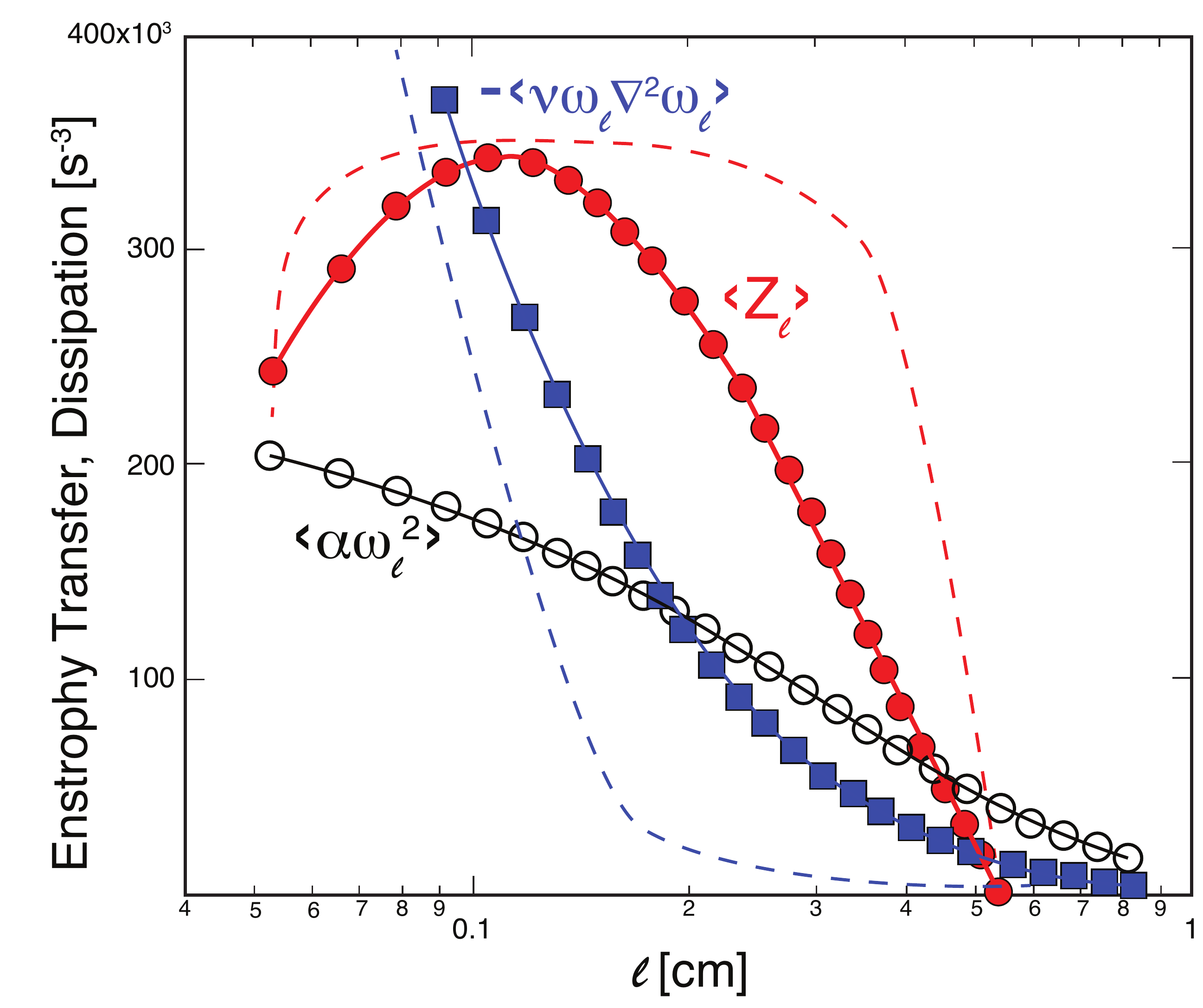}
  \caption{Spatially averaged terms in the coarse-grained enstrophy budget (\ref{largeEnstrophy}) as a function of scale $\ell$ using a Gaussian kernel. Plot shows mean enstrophy flux (dotted red), $\langle Z_\ell\rangle$, which crosses zero at scale $\ell = 5.3$ mm which is approximately the forcing scale. A cartoon schematic (dashed red line) shows the plot that might have been obtained by analyzing the transfer in Fourier space. Plot also shows mean enstrophy dissipation by viscosity (square blue), $\nu\langle|\grad\OL\omega_\ell|^2\rangle$ acting directly at scales $>\ell$ with $\nu = 0.03$ cm$^2$/s. Most dissipation is concentrated at the smallest scales and is negligible at the largest scales. A cartoon schematic (dashed blue line) shows the plot that would have been obtained by analyzing the dissipation in Fourier space.
Open circles (black $\circ$) plot shows mean linear drag due to friction with air, $\alpha\langle\OL\omega_\ell^2 \rangle$ with $\alpha = 5$ sec$^{-1}$. }
   \label{fig:EnstrophyBudgetTerms}
\end{figure}

\begin{figure}[h]
  \includegraphics[width=3in]{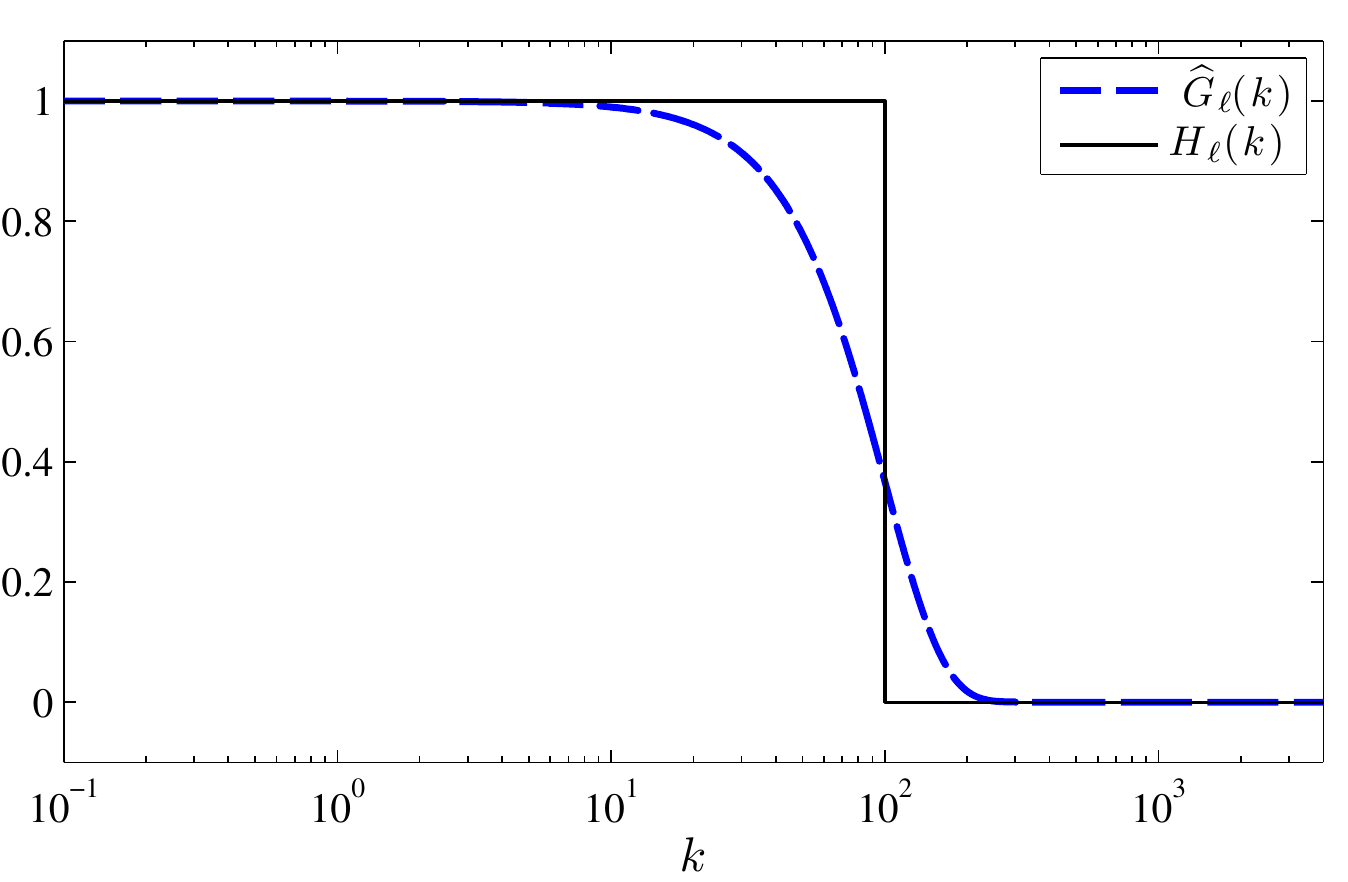}
  \caption{Comparing the sharp spectral, $H_\ell(k)$, and Gaussian, $\widehat{G}_\ell(k)$, filters in Fourier space where $\ell=2\pi/k_\ell$ and $k_\ell=10^2$ in Eq. (\ref{eq:GaussianKernel}). The former has a discontinuous jump whereas the latter is spread out which results in a weighted contribution from more wavenumbers $k$ or, equivalently, scales $\ell$.}
   \label{fig:SharpGaussFilters}
\end{figure}

We now analyze the space average of various terms in the coarse-grained enstrophy budget (\ref{largeEnstrophy}) as a function of scale $\ell$. Figure \ref{fig:EnstrophyBudgetTerms} shows that
the enstrophy flux, $\langle Z_\ell\rangle$, obtained with coarse-graining is smoother as a function of scale $\ell$ relative to the traditional flux obtained in Fourier space (see also Fig. \ref{fig:EnergyFluxSimul}). 
This is because the traditional definition of flux\cite{Frisch_Turbulence} relies on a discontinuous truncation of Fourier modes which corresponds to a sharp-spectral filter 
(or a sinc kernel in x-space, see Eq.(\ref{app_eq:SharpSpectral_x})-(\ref{app_eq:SharpSpectral_k}) along with Fig. \ref{fig:Filters_xk_space} in the appendix) whereas in calculating quantities in Figure \ref{fig:EnstrophyBudgetTerms}, we use a Gaussian kernel, $G_\ell(r)$, that is smooth in Fourier space. As Eq. (\ref{eq:GaussianKernel}) and Fig. \ref{fig:SharpGaussFilters} show, the Fourier transform of a Gaussian, $\widehat{G}_\ell(k)$, is also a Gaussian which is more spread out in
k-space compared to a step function and, thus, entails additional averaging in scale $\ell$ or wavenumber $k$. More generally, following Eyink\cite{Eyink05}, the relation between the traditional flux (in Fourier space), $Z(k)$, and the mean SGS flux, $\langle Z_\ell(\bx)\rangle$, obtained by filtering is 
\be \langle Z_\ell\rangle = \int _0 ^\infty \left(-\frac{d}{dk}|\hat{G}(\ell k/2\pi)|^2\right)Z(k) \, dk,
\lb{filterFourierRelation}\ee
where $\hat{G}(\ell k/2\pi) = \hat{G}_\ell(k)$ is the Fourier transform of $G_\ell(\br)$. Thus, for
a sharp-spectral filter, where $\hat{G}(k)=H_\ell(k)$, 
$$H_\ell(k) =\begin{cases}
    1, & \text{if $|k|<1$}.\\
    0, & \text{otherwise}.\\
  \end{cases}
$$
 the factor $|\hat{G}(\ell k/2\pi)|^2$ in Eq. (\ref{filterFourierRelation}) has a sudden jump to zero and its derivative is a delta function at $k=2\pi/\ell$ which reduces
expression (\ref{filterFourierRelation}) to $Z(k=2\pi/\ell)$. Any filter that 
is spread in k-space, however, will involve more averaging from different scales. Although our analysis
can be done using any kernel, we list the reasons for choosing a Gaussian kernel in Appendix \ref{subsec:DifferentFilters}.
These differences will vanish if we keep increasing the range of scales between enstrophy injection and dissipation
since $\langle Z_\ell \rangle$ becomes constant as a function of $\ell$ and, hence, insensitive to any averaging in scale.
Moreover, our plot of $\langle Z_\ell\rangle$ in Figure \ref{fig:EnstrophyBudgetTerms} is consistent with numerical results reported by Domaradzki \& Carati \cite{DomaradzkiCarati07}, Boffetta\cite{Boffetta07}, and Eyink \& Aluie \cite{EyinkAluie09}. Similar to $\langle Z_\ell\rangle$, the plot of viscous dissipation in Fig. \ref{fig:EnstrophyBudgetTerms} is smoother than a corresponding plot obtained by Fourier analysis. 

As we mentioned earlier, the forcing scale coarsens (grows) with downstream distance because 
of the flow is decaying downstream of the rods. In other words, rather than being forced by vortices shed from the
grid, the turbulence is forced by the mean size of vortices entering the downstream measurement
area.  We identify this effective forcing scale for the enstrophy cascade as the zero crossing point
of the mean filtered enstrophy flux, $\langle Z_\ell\rangle$, as indicated in Figs. \ref{fig:EnergyFlux}, \ref{fig:EnstrophyBudgetTerms}, and \ref{fig:EnsCompare}.

The plot of enstrophy dissipation by linear drag in Fig.\ \ref{fig:EnstrophyBudgetTerms}, $\alpha\langle\OL\omega_\ell^2 \rangle$, is approximately linear suggesting that there is an equal amount of enstrophy dissipation by air drag at all scales. This result is expected from Fig.\ \ref{fig:Spectra} where $\Omega(k)\sim k^{-1}$, which implies that
\begin{eqnarray} 
\alpha\langle\OL\omega^2_\ell\rangle &=&  \const \int _{K_0}^{K} \Omega(k) \,dk = \const \log\left(\frac{L}{\ell}\right) \nonumber\\
&\sim& -\log \left(\ell \right),
\lb{Eq:LinearDrag}
\end{eqnarray}
for $K_0 = 2\pi/L$ and $K=2\pi/\ell$. Eq. (\ref{Eq:LinearDrag}) is consistent with a linear plot of $\alpha\langle\OL\omega^2_\ell\rangle$ on a log-linear graph. More precisely, the downward curvature of $\alpha\langle\OL\omega^2_\ell\rangle$ is consistent with a slightly steeper dependence $\Omega(k) \sim k^{-(1+a)}$ with $a \approx 0.3$ (corresponding to $\beta \approx 3.3$ in $E(k)$).  This value of $a$ is consistent with the spectra $\Omega(k)$ in Fig.\ \ref{fig:Spectra} where the slopes are slightly steeper than $k^{-1}$.  They are also consistent with single point measurements of energy spectra in soap films \cite{MartinPRL98}. 

\subsection{Relation to third-order structure functions}

\begin{figure}[h]
  \includegraphics[width=3.5in]{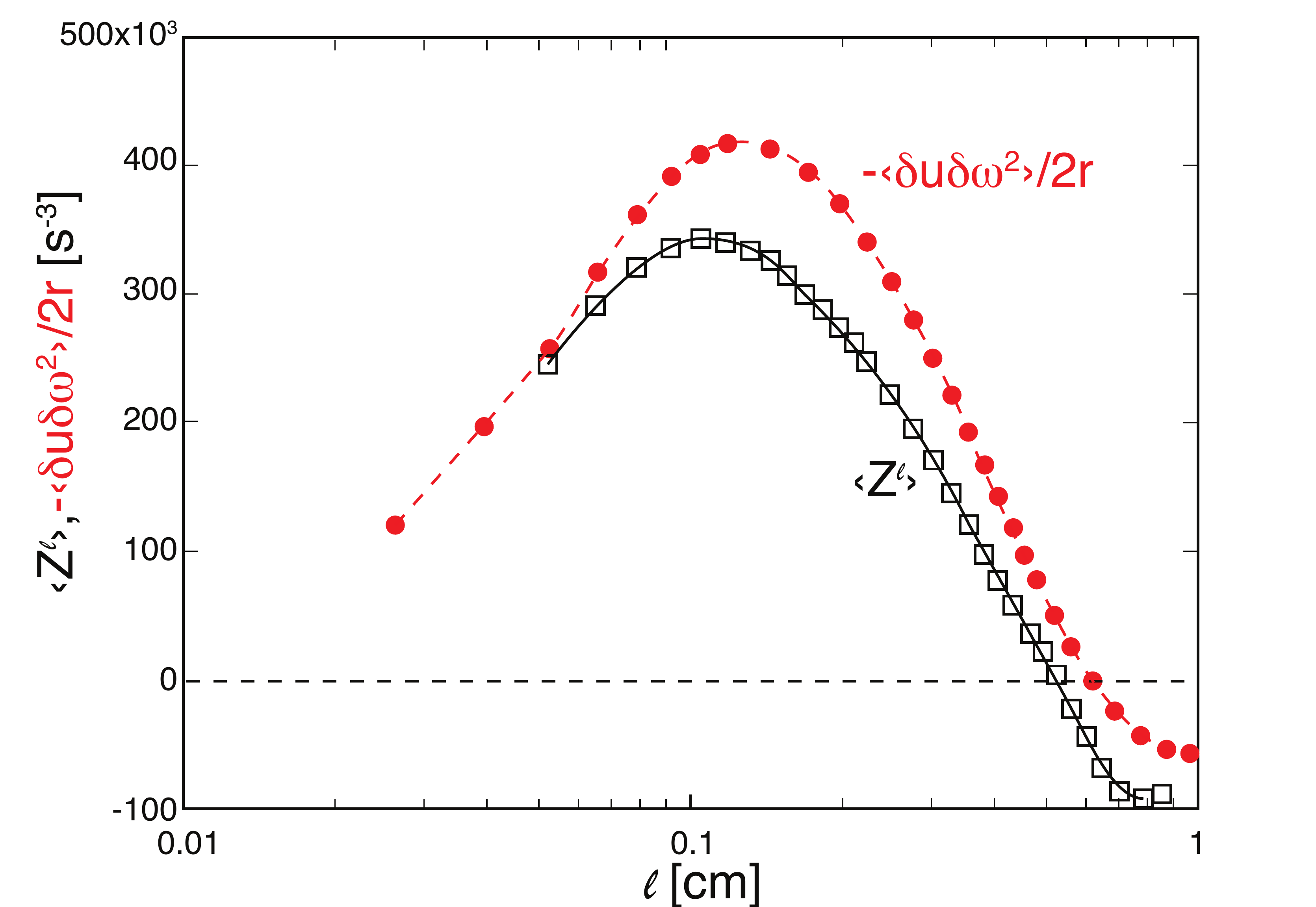}
  \caption{Comparing mean enstrophy flux, $\langle Z_\ell\rangle$, (open black square) with a more traditional measure using third-order structure function, $SW_2(\ell)$ (solid red circle). As discuss in the text, $r = 2\ell$. The two plots show very good agreement, within factor $\approx 3/4$, as shown analytically\cite{Eyink95b}.}
   \label{fig:EnsCompare}
\end{figure}
A result analogous to Kolmogorov's $4/5$th law of incompressible 3D turbulence was shown to exist for 2D flows by Polyakov\cite{Polyakov93} and Eyink\cite{Eyink95b}. It describes the enstrophy cascade:
\be SW_2(r)=\langle\delta u_\parallel(r)\, |\delta \omega(r)|^2 \rangle = -2 \,r\, \eta,
\lb{EnstrophyStructure}\ee
where $\eta$ is the viscous dissipation rate of enstrophy, $\delta u_\parallel(\br) = [\bu(\bx+\br)-\bu(\bx)]\bdot\hat{\br}$ is a longitudinal velocity increment, and $r = |\br|$. It should be noted that relation (\ref{EnstrophyStructure}) is derived under the assumption of an isotropic homogeneous flow at inertial scales over which there is negligible effects from viscosity, boundaries, forcing, and air drag. 

What is probably less appreciated in the literature is that there is a direct relation, due to Eyink \cite{Eyink95b}, between the enstrophy flux, $\langle Z_\ell\rangle$, obtained by coarse-graining and the more traditional third-order structure function, $SW_2(r)$,
\be \langle Z_\ell\rangle = -\langle  \grad\OL\omega_\ell \bdot \OL\tau(\bu,\omega) \rangle
= \mathcal{O}\left(\frac{\delta u (\ell)\,|\delta \omega(\ell)|^2}{\ell}\right).
\lb{ZSWrelation}\ee
Relation (\ref{ZSWrelation}) implies that the two measures, $SW_2$ and $\langle Z_\ell\rangle$, should agree within a factor of order unity and should have the same scaling as a function of $\ell$ (or $r$). We note that applying a Gaussian filter at point $\bx$ averages the flow within a circle of radius $\ell/2$ (see Fig. \ref{fig:Filters_xk_space} in the appendix) whereas the structure function, $SW_2(\bx; r)=\delta u_\parallel(\bx;r)\, |\delta \omega(\bx;r)|^2 $, uses information within a circle of radius $r=|\br|$
around $\bx$. Hence, to compare $\langle Z_\ell\rangle$ and $SW_2(r)$ in Fig. \ref{EnstrophyStructure}, we set $r=2\ell$.
Figure \ref{fig:EnsCompare} shows that this is indeed the case in our experiment, where we find that $\langle Z_\ell\rangle$ and $SW_2(r)$ are almost equal at all scales $\ell$ we measured, within a factor $\approx 3/4$.

\subsection{Spatial Statistics of the Cascade\lb{sec:SpatialStatistics}}
\begin{figure}
  \includegraphics[width=3in]{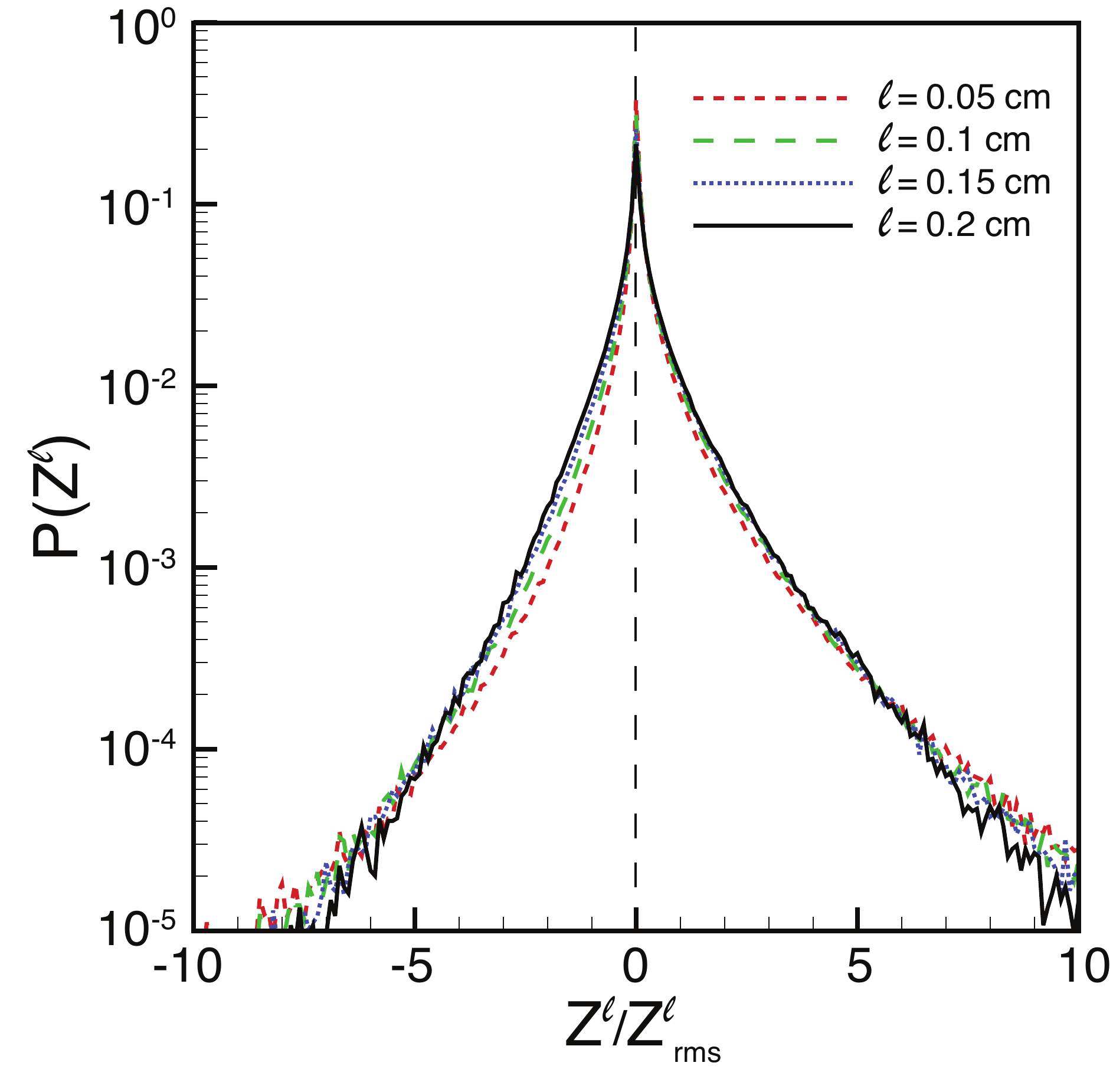}
  \caption{Spatial distribution of the enstrophy flux, $Z_\ell(\bx)$, across different scales $\ell$. Positive values, covering $60\%$ of the domain's area, correspond to locations where enstrophy is transferred from scales larger than $\ell$ to scales smaller than $\ell$. Negative values, covering $40\%$ of the domain's area, correspond to points with backscatter from small to large scales. The pdf has a positive mean, indicating that there is a net transfer of enstrophy to smaller scales, $\langle Z_\ell\rangle > 0$, consistent with
Fig. \ref{fig:EnstrophyBudgetTerms}. }  
   \label{fig:Zpdf} 
\end{figure}

Although we observe a positive mean flux of enstrophy to small scales, $\langle Z_\ell\rangle > 0$, the spatially averaged quantity does not contain any information about the scale-dynamics at various locations in space. The coarse-graining approach allows us to resolve the scale-dynamics in space. As we showed qualitatively in Figures \ref{fig:enstrophy-flux-pictorial}, the space-local enstrophy flux field, $Z_\ell(\bx)$, is inhomogeneous with distinct spatial characteristics such as 
(i) the presence of both upscale and downscale transfer of enstrophy at different locations and that the net positive rate of a downscale cascade, $\langle Z_\ell\rangle > 0$, is only a result of cancellations between forward-scatter and back-scatter, and (ii) the enstrophy flux field has small patches of intense positive and negative values relative to the spatial mean, {\it i.e.} it is intermittent.

Figure \ref{fig:Zpdf} plots the probability density function (pdf) of the field $Z_\ell (\bx)$ for different values of scale $\ell$. The pdf has heavy tails, confirming the qualitative observations from Figure \ref{fig:enstrophy-flux-pictorial} of an intermittent field. We observe that the positive mean, $\langle Z_\ell\rangle > 0$, is a result of strong cancellations between negative and positive values, consistent with previous results\cite{Chenetal03,Boffetta07}. 

From Fig. \ref{fig:Zpdf}, we find that the nonlinear dynamics transfers enstrophy downscale with a $60\%$ probability and upscale with a $40\%$ probability. Our finding gives empirical support to theoretical results due to Merilees \& Warn\cite{MerileesWarn75}. The authors in Ref.\cite{MerileesWarn75} used basic conservation laws, following the earlier work of Fjortoft\cite{Fjortoft53}, along with a counting argument to infer that the number of Fourier wavevector triads in a 2D flow transferring enstrophy to smaller scales is greater than the number of triads transferring enstrophy to larger scales, with a ratio of $60\%$ to $40\%$. Their derivation does not require the existence of an intertial range and is, hence, applicable to any 2D flow. However, as remarked by the authors themselves\cite{MerileesWarn75} and several others\cite{Kraichnan67,TungWelch01,GkioulekasTung07,BurgessShepherd13},
such counting arguments are kinematic in nature and do not guarantee {\it a priori} that the triads being counted will participate in the dynamics and in the transfer of enstrophy across scales. Moreover, their counting argument does not predict if all triads transfer enstrophy equally or if some triadic interactions should be weighted more heavily than others.

Our plot in Fig. \ref{fig:Zpdf} suggests that indeed, the kinematic arguments of Merilees \& Warn\cite{MerileesWarn75} are dynamically realized in a 2D turbulent flow. We speculate that this is due to the ergodic nature of a turbulent flow, which allows the dynamics to sample the entire phase-space with equal probability, thereby allowing all kinematically possible triads to be dynamically active and contribute equally to enstrophy transfer. We also speculate that in quasi-laminar 2D flows, which are not ergodic, the predictions of Merilees \& Warn\cite{MerileesWarn75} will not hold true. In fact, we shall see in section \ref{sec:EnstrophyFluxFlowTopology} below that when we restrict our analysis to coherent structures in the flow, the $60\%-40\%$ result breaks down.

The $60\%$-$40\%$ result has been previously reported from numerical data\cite{Chenetal03}, and to the best of our knowledge, Fig. \ref{fig:Zpdf} is the first experimental evidence of Merilees \& Warn's prediction. It is worth noting that measuring individual triadic interactions in a 2D flow to directly check the theoretical results\cite{MerileesWarn75}  would require ${\mathcal O}(N^4)$
evaluations, where $N\times N$ is the number of grid points used in interpolating the velocity data. This is prohibitively expensive for 
any well-resolved flow with a significant scale-range to capture the cascade. Our $60\%$ -- $40\%$ result was made possible by the coarse-graining approach, which allowed us to measure upscale and downscale transfers, not in individual triads, but at each
position $\bx$.

\subsection{Enstrophy Flux and Flow Topology\lb{sec:EnstrophyFluxFlowTopology}}
As mentioned earlier, the physical mechanism responsible for the enstrophy cascade is vortex gradient stretching from mutual-interaction among vortices\cite{BoffettaEcke12}. To better understand the physics behind results presented in the previous section, it is important to investigate the dynamical role of vortices and their influence on the cascade. To determine vorticity regions,  we use an Eulerian coherent structure identification technique, which we dub `BPHK' after work by
Basdevant \& Philipovitch\cite{BasdevantPhilipovitch94} and Hua \& Klein\cite{HuaKlein98}.
The BPHK method detects coherent structure regions in the flow using a constrained version of the 
Okubo-Weiss criterion\cite{Okubo70,Weiss91}.

\subsubsection{BPHK and the Okubo-Weiss criterion}
Following the presentation given in Ref.\cite{BasdevantPhilipovitch94}, 
we start with a brief description of Okubo-Weiss. Assuming that we have an inviscid 2D fluid, the time evolution of the vorticity field is given by
$$ \frac{d}{dt}\omega = \partial_t \omega + (\bu\bdot\grad)\omega = 0
$$
which implies that vorticity following a fluid parcel is conserved in an inviscid flow. It follows that the evolution of vorticity gradient, $\grad\omega$, along a fluid parcel is:
\be \frac{d}{dt}\partial_i\omega + (\partial_i u_j)\,\partial_j\omega = 0.
\lb{GradVort}\ee
If $\grad\bu$ in Eq. (\ref{GradVort}) evolves slowly compared to the vorticity gradient $\grad\omega$, then $\grad\bu$ is essentially constant and the evolution of $\grad\omega$ is fully determined by the velocity gradient.
Since $\grad\bu$ is traceless by incompressibility, the local topology is determined by its determinant, $\mbox{det}(\grad\bu)$. In the case where $\mbox{det}(\grad\bu)> 0$, the eigenvalues of $\grad\bu$ are imaginary and the vorticity gradient is subject to a rotation. Otherwise, the eigenvalues are real and the vorticity gradient is subject to a strain, which enhances the vorticity gradient. Thus, according to the Okubo-Weiss approximation, there are two types of topological structure within a flow: centers for which $\mbox{det}(\grad\bu)\gg 0$ and saddles for which $\mbox{det}(\grad\bu)\ll 0$.

The critical assumption in Okubo-Weiss, as pointed out by Basdevant \& Philipovitch\cite{BasdevantPhilipovitch94}, is that $\grad\bu $ evolves slowly compared to $\grad\omega$. The assumption can be made more quantitative by taking the material derivative of Eq. (\ref{GradVort}),
$$ \frac{d^2}{dt^2}\partial_i\omega + \partial_j\omega \frac{d}{dt}(\partial_i u_j) + (\partial_i u_j)\frac{d}{dt}(\partial_j\omega)=0.
$$
This relation makes it clear that Okubo-Weiss rests on assuming
$$ \frac{|\grad\omega \bdot \frac{d}{dt}(\grad \bu)|}{|(\grad \bu)\frac{d}{dt}(\grad\omega)|} \ll 1.
$$
The inequality can be rewritten\cite{BasdevantPhilipovitch94,HuaKlein98} in terms of the 
eigenvalues, $\lambda_\pm$, of the pressure Hessian, $H_{ij} = \partial_i\partial_j p$, yielding
\be R\equiv \frac{(\lambda_+-\lambda_-)^2}{(\lambda_+ + \lambda_-)^2} \ll 1,
\lb{eq:BPHKcondition}\ee
as a validity condition on the Okubo-Weiss criterion. The implementation of BPHK from experimental data is fairly straightforward if one has high spatial resolution of the velocity field. From the measured velocities, we evaluate\cite{RiveraWu00} the pressure field by solving $\nabla^2p = 2\mbox{det}(\grad\bu)$.
Then the field $R$ is obtained from the pressure Hessian. Spatial positions where $R < 1$ correspond to places where the Okubo-Weiss assumption is valid. For those positions, regions for which $\mbox{det}(\grad\bu)<0$ are identified as saddles and those with $\mbox{det}(\grad\bu) > 0$ are centers. These points comprise the coherent structures as shown in Fig. \ref{Fig:BPHK}.
\begin{figure}[h!]
   \includegraphics[height=4in,width=2in]{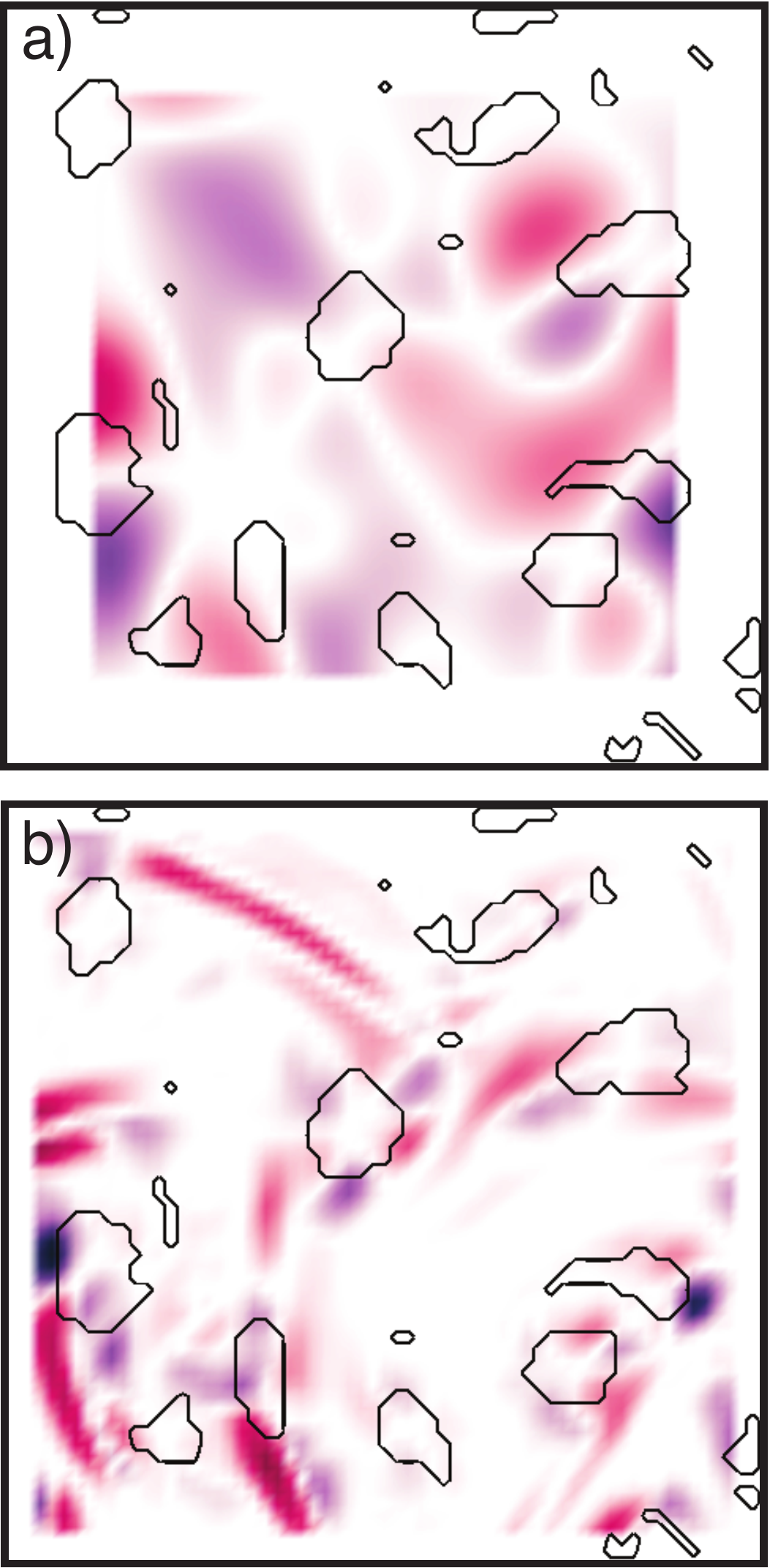}
  \caption{Applying the BPHK to our flow. Contours indicate the coherent structures detected.
Colors indicate positive (red) and negative (blue) enstrophy flux, $Z_\ell(\bx)$, at scales 
  (a) $\ell=0.2$ cm and (b) $\ell=0.05$ cm.
     }
\lb{Fig:BPHK}  \end{figure}

\subsubsection{Flow Topology and the Enstrophy Cascade}
Figure \ref{PDF_CenterSaddle} shows the probability density function (pdf) 
of the flux, $Z_\ell$, in these coherent structure regions. We find that centers, 
associated with elliptic regions in the flow such as vortex cores,
are very inefficient at transferring enstrophy to smaller scale. We find that the dynamics
inside centers leads to an upscale and downscale enstrophy cascade with almost
equal probability. More precisely, across any scale $\ell$, the flow inside centers transfers
enstrophy to scales $<\ell$ with a $55\%$ probability and to scales $>\ell$ with a $45\%$ probability.

\begin{figure}[h!]
  \includegraphics[width=3.5in]{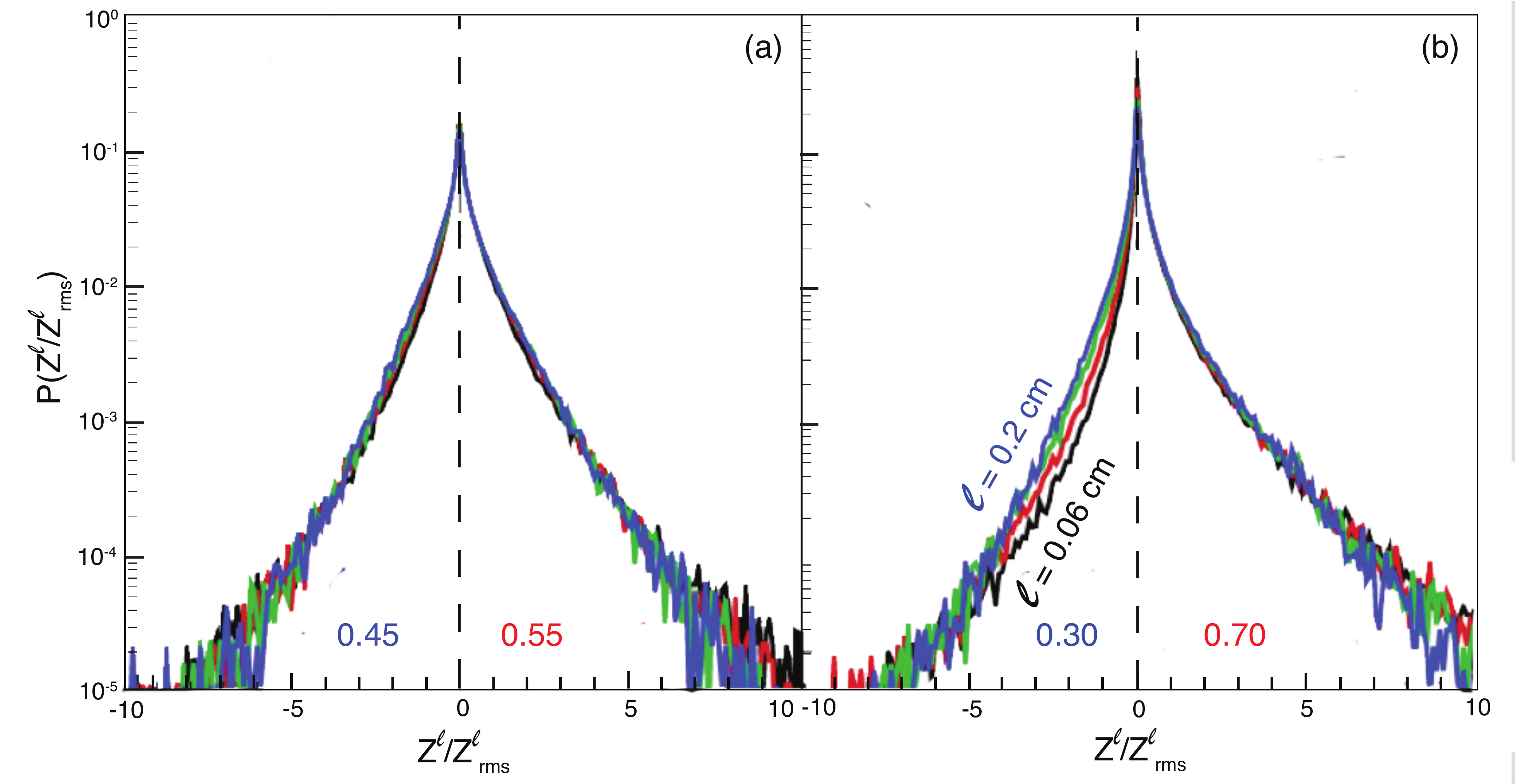}
  \includegraphics[width=2.5in]{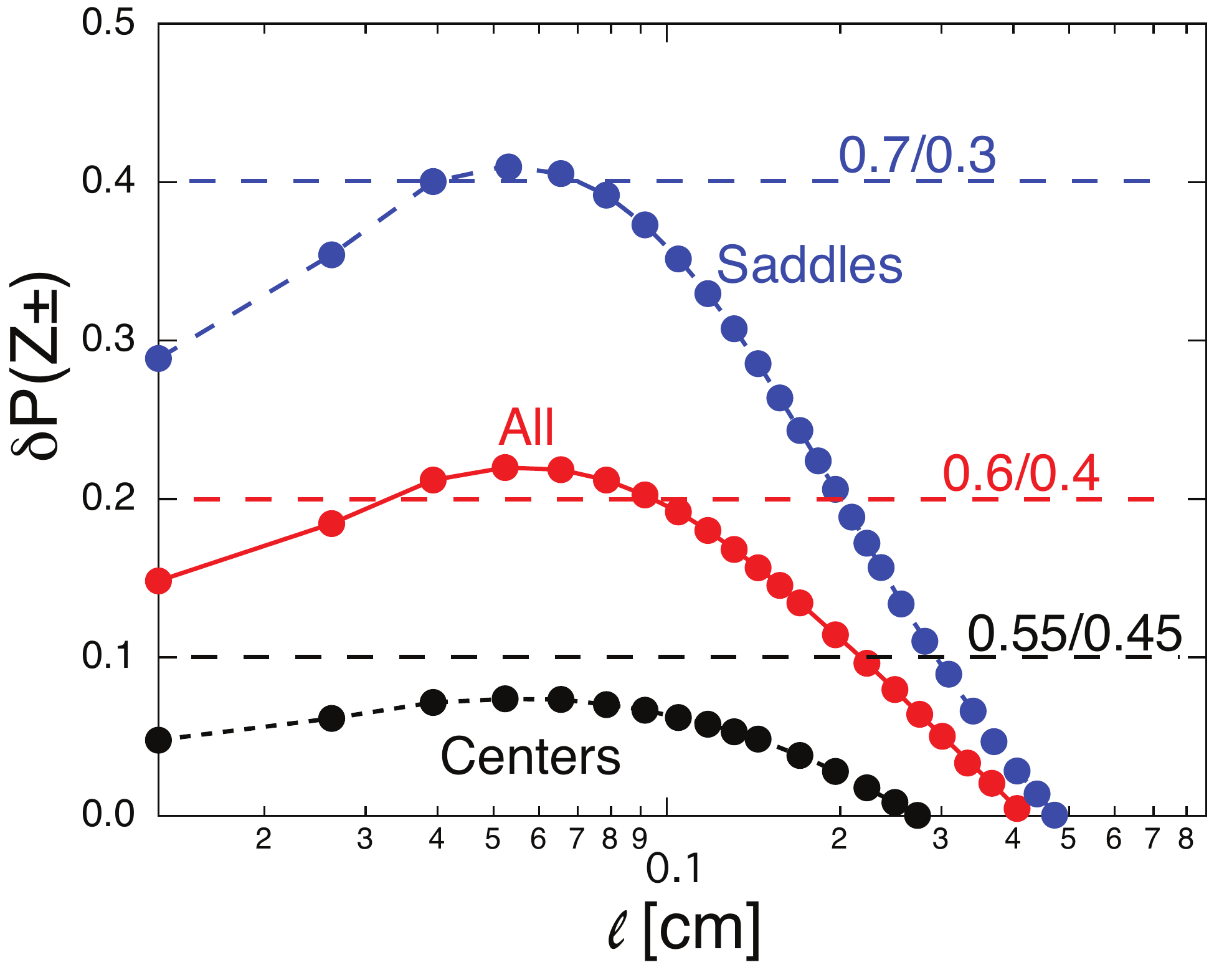}
  \caption{
Distribution function of the enstrophy flux, $Z_\ell(\bx)$, across different scales $\ell$ within coherent structures identified by the BPHK method. Panel (a) shows that centers have considerable backscatter and transfer enstrophy to small and large scales with almost equal probability. Panel (b) shows that saddles are relatively more efficient at transferring enstrophy to smaller scales with a $70:30$ ratio.
In panel (b), the top curve (blue) is for $\ell=0.2$ cm, the bottom (black) curve is for $0.06$ cm, and the in-between curves correspond to (red) $\ell=0.1$ cm and (green) $\ell=0.15$ cm.  Note that the curves in panel (a) correspond to the same set of scales $\ell$ but are not distinguishable. Panel (c) shows the difference in probability of positive and negative flux values, $\delta P(Z_\ell) = P^+_\ell-P^-_\ell$, as a function of scale $\ell$. It shows that in centers (black dotted line), enstrophy is transferred with almost equal probability, $P^+_\ell \approx P^-_\ell$, to small and to large scales. Centers are less efficient at cascading enstrophy compared to the domain average (red solid line). On the other hand, saddles (blue dashed line) are more efficient than the domain mean over the entire range of scales.}
\lb{PDF_CenterSaddle}
\end{figure}

In contrast, we find from Fig. \ref{PDF_CenterSaddle} that saddles, associated with 
hyperbolic regions of large strain, are much more efficient at cascading enstrophy to smaller scales.
The dynamics inside saddle regions cascades enstrophy downscale twice as often as upscale.
More precisely, across any scale $\ell$, the flow inside saddle regions transfers
enstrophy to scales $<\ell$ with a $70\%$ probability and to scales $>\ell$ with a $30\%$ probability.

In Fig. \ref{PDF_CenterSaddle}, we also plot the difference in probability for a downscale, $P^+_\ell=P(Z_\ell>0)$,
and an upscale, $P^-_\ell=P(Z_\ell<0)$, cascade. As a function of $\ell$, the plot of 
$\delta P(Z_\ell^\pm) = P^+_\ell - P^-_\ell$, in Fig. \ref{PDF_CenterSaddle} shows that centers 
are inefficient at transferring enstrophy over the entire cascade range compared to the domain average. 
In contrast, saddle regions are systematically more efficient at carrying enstrophy to small scales
compared to the domain average, for the entire range of scales. Our findings show that 
the $60\%-40\%$ result of Merilees \& Warn\cite{MerileesWarn75} does not hold in coherent structures but only
in an average sense over the entire turbulent flow domain. As mentioned in section \ref{sec:SpatialStatistics},
we speculate that the $60\%-40\%$ result would only hold in turbulent flows because they are ergodic and ensure that all triads participate (equally) in the enstrophy transfer, as required by the counting argument of 
Merilees \& Warn. We expect the $60\%-40\%$ result to break down in quasi-laminar 2D flows, such as in coherent structures
analyzed in Fig. \ref{PDF_CenterSaddle}, because they are not ergodic.

The increased efficiency of saddle regions relative to centers is to be physically expected.
Centers can be qualitatively characterized as regions of rotation in the flow where little amplification
of gradients occurs. Saddles, on the other hand, are straining regions where vorticity iso-contours
are stretched and crammed together yielding an amplification of vorticity gradients, which is equivalent
to the generation of small-scales. The coarse-graining approach, which affords the simultaneous 
resolution of the nonlinear dynamics both in space and in scale, along with structure identification 
tools enabled us to quantify such qualitative behavior.

\subsection{Energy Transfer}

\begin{figure}[h]
   \includegraphics[height=2.5in,width=2.5in]{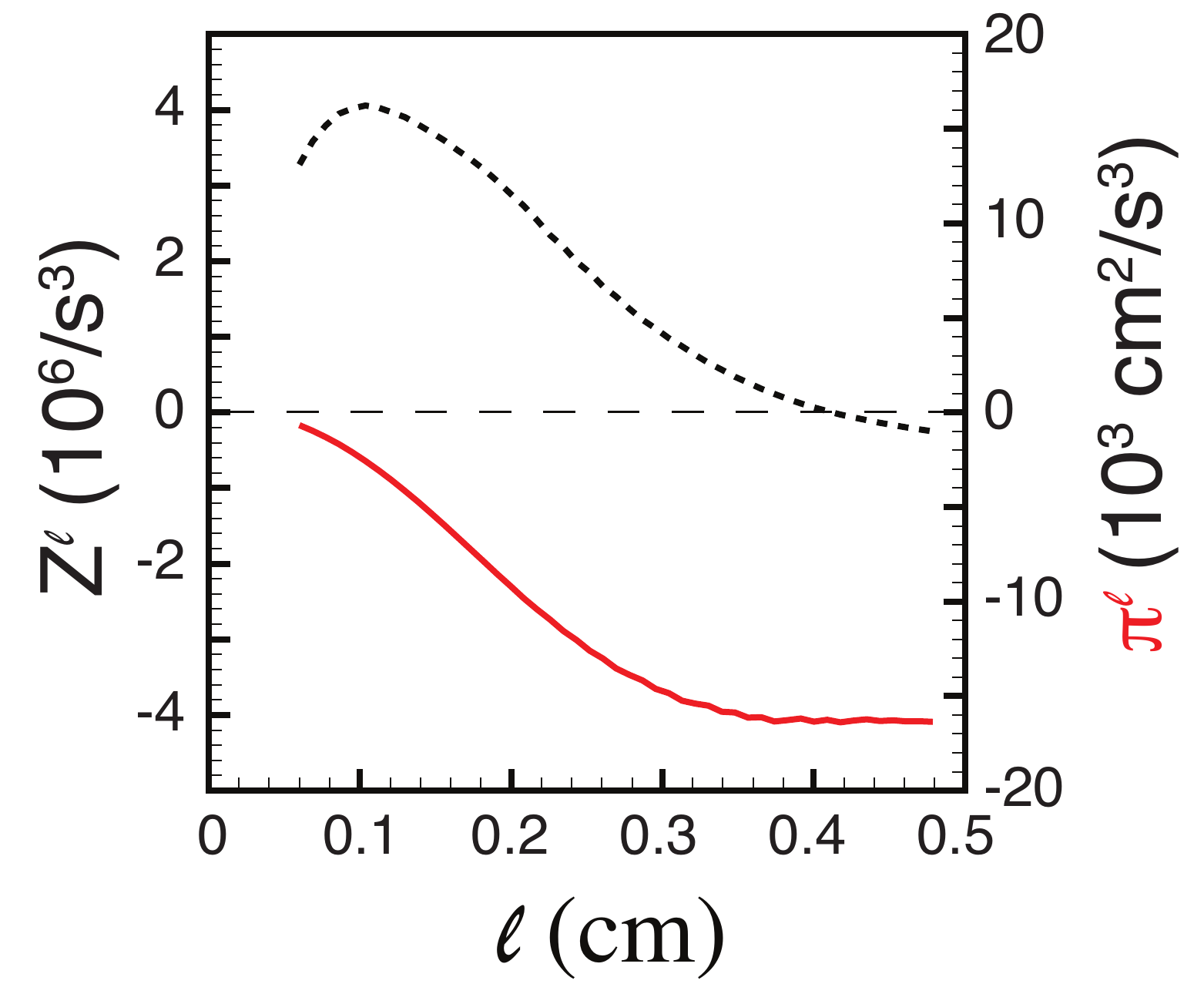}
   \includegraphics[height=2.5in,width=2.5in]{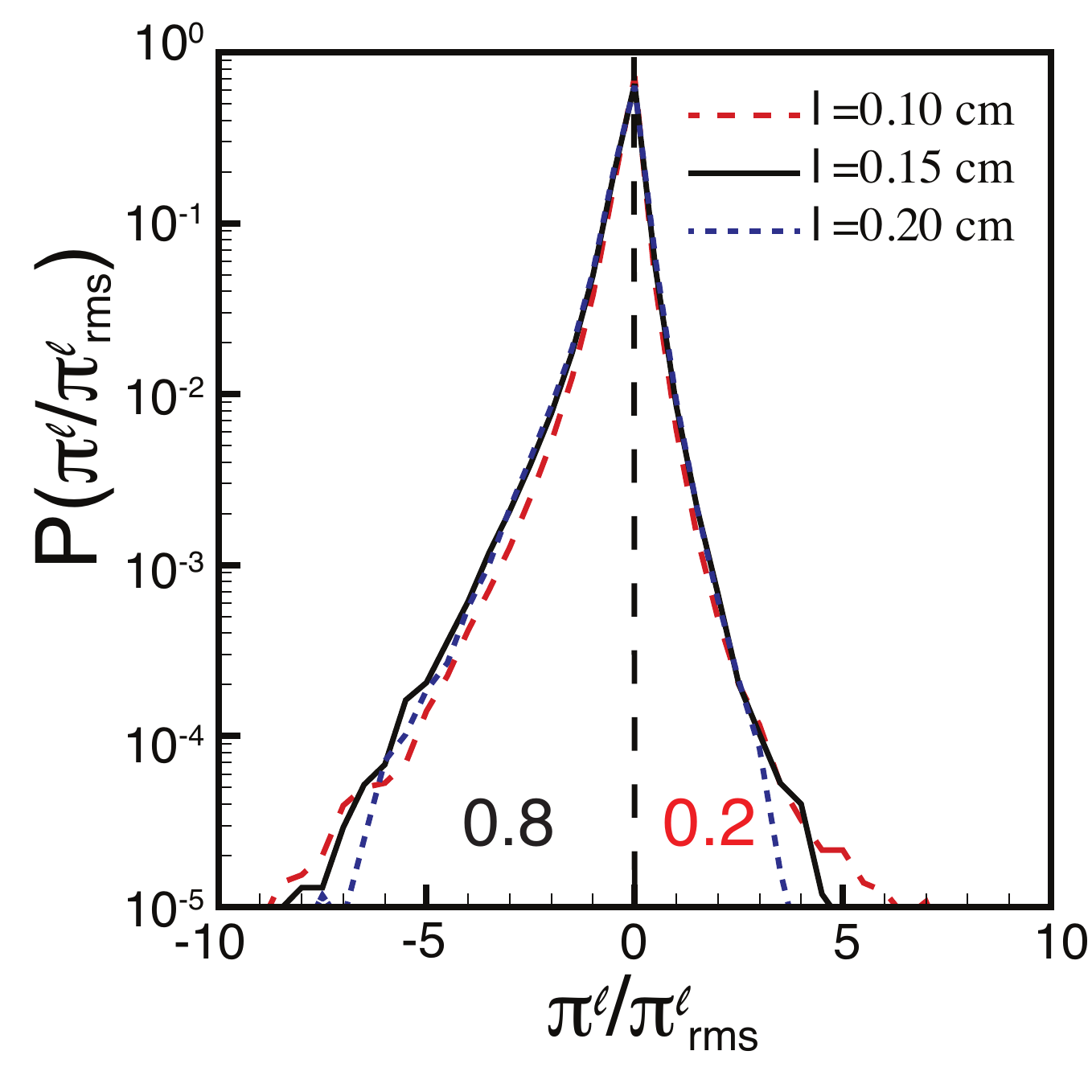}
   \caption{Left panel shows that mean energy flux (solid red plot), $\langle\Pi_\ell\rangle$, is negative over a wide 
   range of scales $\ell$ in our experiment ($\langle\Pi_\ell\rangle$ decays to zero at scales larger than those shown). 
   We also show the mean enstrophy flux (dashed black plot), $\langle Z_\ell\rangle$,
   which is positive over most scales analyzed. These data are for a mean location of 5 cm downstream as opposed
to the 6 cm data shown below.  Thus, the magnitude of $\langle Z^\ell \rangle$ is larger and the zero crossing point is slightly smaller (see next subsection for details).
   Right panel shows that the the probability density function (pdf) of $\Pi_\ell(\bx)$ is negatively skewed 
for scales $\ell = 0.1$, $0.15$, $0.2$ cm. This indicates an upscale energy transfer over 80\% of the domain.
   \label{fig:EnergyFlux} }
\end{figure}

It is unclear from existing 2D turbulence phenomenologies how energy 
should be transferred between scales in our soap film apparatus. 
The flow we are analyzing is statistically 
stationary but decaying downstream of the rods. The most energetic vortices in the 
flow coarsen (grow in size) downstream of the rods
(which can be seen in Fig. \ref{fig:Spectra} as a shift in the peak of $E(k)$ to smaller $k$
and in the contour plot of Fig. 16 in Ref.\cite{Vorobieffetal99}) and, 
hence, the effective forcing scale becomes larger than that of the rod separation. We 
identify the effective forcing scale as the zero-crossing point of $\langle Z_\ell\rangle$, 
which also shifts to larger scales (compare plots of $\langle Z_\ell\rangle$ 5 cm downstream in Fig. \ref{fig:EnergyFlux} 
and 6 cm downstream in Fig. \ref{fig:EnstrophyBudgetTerms} ).
The range of scales we are probing is
that of downscale enstrophy cascade over which a simultaneous energy cascade (with a scale-independent energy flux)
is not expected from theory\cite{Fjortoft53,Kraichnan67,Leith68,Batchelor69}.

\begin{figure*}[!htbp]
   \includegraphics[height=3in,width=3in]{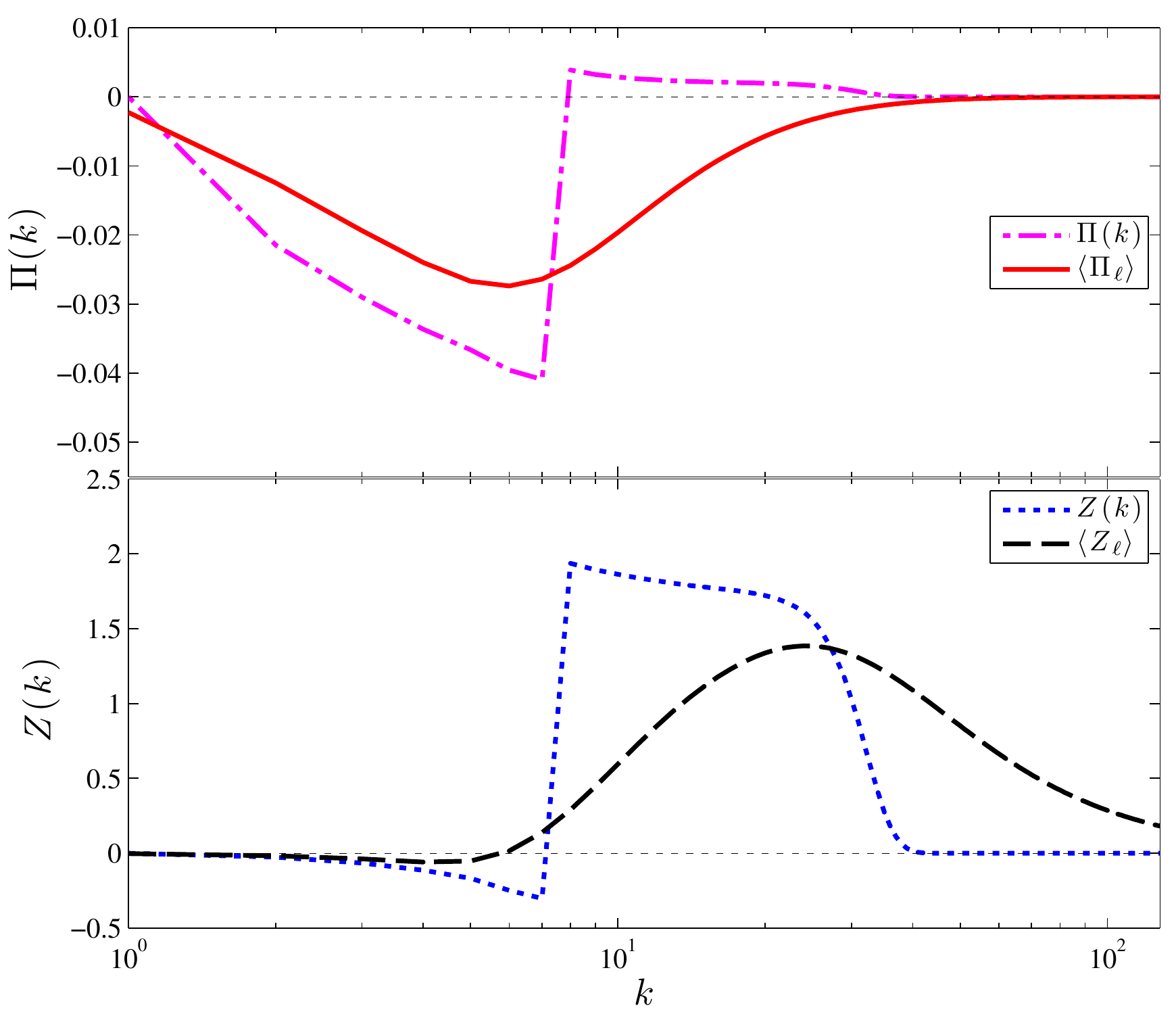}
   \includegraphics[height=3in,width=3.3in]{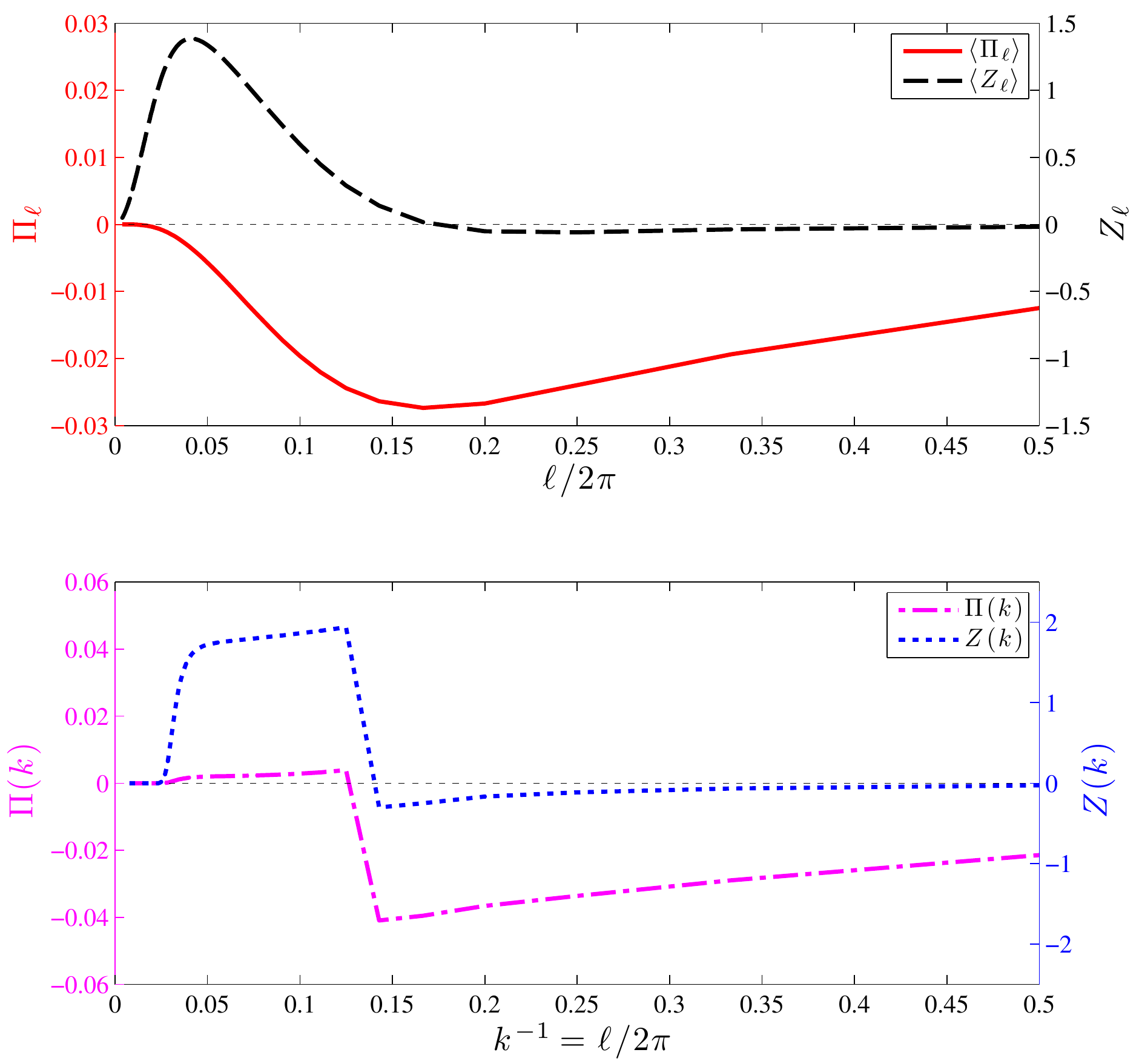}
   \caption{
   Plots of energy and enstrophy fluxes from a numerical simulation to compare with Fig. \ref{fig:EnergyFlux}.
   Forcing was at $k_f=7$.
   Upper-right panel is very similar to that from experiment in Fig. \ref{fig:EnergyFlux}:  it shows mean energy 
   flux (solid red), $\langle\Pi_\ell\rangle$, and mean enstrophy flux (dashed black), $\langle Z_\ell\rangle$, obtained 
   by coarse-graining with a gaussian kernel.
   Lower-right panel shows the traditional energy (dashed-dotted magenta) and enstrophy (dotted blue) spectral fluxes obtained by
   a sharp-spectral filter in Fourier-space, plotted here as a function of $k^{-1}$.
   Left two panels plot the fluxes as a function of Fourier wavenumber, $k$, on log-linear axes. Upper-left panel shows the traditional 
   energy flux, $\Pi(k)$, (dashed-dotted magenta) and that obtained by coarse-graining, $\langle\Pi_\ell\rangle$, (solid red).
   Lower-left panel shows the traditional enstrophy flux, $Z(k)$, (dotted blue) and that obtained by coarse-graining, 
   $\langle Z_\ell\rangle$ (dashed black). Note that fluxes computed by coarse-graining are smoother than 
   (and can be derived from) their Fourier spectral counterparts (see text and Eq. (\ref{filterFourierRelation})).
}   \label{fig:EnergyFluxSimul}
\end{figure*}

Nevertheless, it is interesting
to measure energy transfer in this system which may reflect the behavior observed more generally in systems with quasi-2D character.  In Fig. \ref{fig:EnergyFlux}, we see that the mean energy transfer is negative revealing that energy is being transferred upscale despite the expected lack of an energy cascade ({\it i.e.}, a constant energy flux). In order to better understand such a plot, it is worth comparing to Fig. \ref{fig:EnergyFluxSimul} where we also show the more traditional spectral fluxes computed in Fourier space. As we explain in more detail in section \ref{sec:SpaceAveragedTerms} and Eq. (\ref{filterFourierRelation}), fluxes obtained by coarse-graining are smoother functions of scale (or wavenumber) because they involve more averaging in scale compared to spectral fluxes.

Fig. \ref{fig:EnergyFlux} shows the spatial distribution (pdf) of energy transfer, $\Pi_\ell(\bx)$. We find that regions where energy is transferred from scales $<\ell$ to larger scales ($\Pi_\ell(\bx) < 0$) occupy $80\%$ of the domain for several values of $\ell$. Indeed, this behavior is reflected in the spectra of Fig. \ref{fig:Spectra}, where although energy is decaying at all scales downstream of the grid, the spectrum maximum shifts to smaller $k$ indicating that there is a slight relative build-up of energy at those large scales. 

Nevertheless, the energy spectra scale as $E(k)\approx k^{-3}$ and show no signature of a sustained inverse energy {\it cascade}. This is consistent with the Kraichnan-Leith-Batchelor theory \cite{Kraichnan67,Leith68,Batchelor69} where we know that concurrent cascades of energy and enstrophy, which would necessitate that energy and enstrophy fluxes be scale independent, $\langle\Pi_\ell\rangle = \const$ and $\langle Z_\ell\rangle = \const$, is prohibited.   It is important to differentiate between an inverse energy {\it transfer} that happens generically in 2D flows from inverse energy {\it cascade} which requires a \emph{constant} negative energy flux.
By definition, a cascade should persist to as small (or large) scales as are available, regardless of how small (or large) the viscous scale (or drag scale) is.  In systems of limited size such as laboratory experiments or numerical simulations, one would like to see a decade of approximately constant flux to definitively demonstrate a cascade.  Transfer, on the other hand, is less restrictive. There can be a significant \emph{transfer} of energy in a system of limited scale-range but such \emph{transfer} would vanish once enough scale-range has been established (e.g., a high Reynolds number flow). Such transfer is a finite $Re$ number effect and will not survive extrapolation from an experimental system of limited scale-range to natural systems characterized by many of decades of wave numbers. For example, in Fig. \ref{fig:EnergyFluxSimul} (lower-left panel) the enstrophy flux, $Z(k)<0$ over a small range of wavenumbers, indicating a transfer of enstrophy to larger scales. Such transfer is, however, not necessarily indicative of a \emph{cascade} because it is not persistent and cannot carry enstrophy to arbitrarily large scales. Similarly (upper-left panel of Fig. \ref{fig:EnergyFluxSimul} ) for the energy flux, $\Pi(k)>0$ over a limited range of scales cannot be considered to be a cascade of energy to small scales. We know that if viscosity in such a flow is made smaller, energy would not persistently get transferred to arbitrarily small viscous scales. Hence, this $\Pi>0$ region is energy \emph{transfer} to small scales but not a \emph{cascade}. The take home message here is that the existence of transfer is a necessary but not sufficient condition for demonstrating a cascade, a confusion that persists in the literature.

\section{Conclusions \label{sec:Conclusions}}
In this paper, we carried out an experiment of 2-dimensional turbulence in soap film
to probe the enstrophy cascade locally in space and correlate it with topological features in the flow. 
Our investigation was made possible by a coarse-graining technique
\cite{Leonard74,Germano92,Eyink95a,Eyink95b}
rooted in the subjects of mathematical analysis of PDEs and LES turbulence modeling, that 
allows the analysis of nonlinear dynamics and scale-coupling simultaneously in scale and in space.

We analyzed the behavior of various terms in the coarse-grained enstrophy
budget over the range of scales resolved by our measurements. We also verified 
an exact relation due to Eyink \cite{Eyink95b} relating traditional third-order structure function
to the enstrophy flux. The coarse-graining method allowed us to also calculate the transfer of 
enstrophy across scale  $\ell$ at every point $\bx$ in the flow domain. 

By resolving the enstrophy flux in space, we were able to measure the probability 
for the nonlinear dynamics to transfer enstrophy
upscale versus downscale. We found that enstrophy cascades to smaller (larger) scales 
with a 60\% (40\%) probability, in support of theoretical predictions by Merilees \& Warn\cite{MerileesWarn75}.
The analysis in Ref.\cite{MerileesWarn75} is purely kinematic and is not guaranteed {\it a priori} to be satisfied
by the actual dynamics. We speculated that their predictions are valid in our soap film flow due to the ergodic nature of turbulence, and conjectured that in quasi-laminar 2D flows, which are not ergodic, the predictions of Ref.\cite{MerileesWarn75} will break down. 

Our conjecture has some support from an analysis we did of the enstrophy cascade in coherent structures in 
section \ref{sec:EnstrophyFluxFlowTopology}, where the $60\%-40\%$ result breaks down.
We identified coherent structures using the so-called BPHK technique\cite{BasdevantPhilipovitch94,HuaKlein98} 
to detect ``centers'' (often associated with vortex cores) and ``saddles'' 
(often associated with strain regions). We were able to gain insight into the mechanism
behind the cascade by spatially correlating these topological features with enstrophy transfer. 
We found that centers are very inefficient at transferring enstrophy between scales, in contrast to saddle 
regions which transfer enstrophy to small scales with a $70\%$ probability. 

The coarse-graining framework assumes very little about the nature of turbulence, making it
a powerful technique to study complex non-canonical flows for which our theoretical understanding may be rudimentary.
Standard tools that have been developed and used in the study of turbulence are often only valid for homogeneous isotropic incompressible flows. The approach we utilized here is very general and is not restricted by these requirements ---it does not even require the presence of turbulence, as long as there are multiple scales interacting nonlinearly with each other. 

We believe that the applicability of this technique to numerical as well as experimental data holds great promise to
provide insight into nonlinear instabilities, mixing, and turbulence in 2D and 3D flows.

\section{acknowledgments}
We thank M. Chertkov, G. Eyink, P. Marcus, T. Shepherd, and B. Shraiman for
interesting discussions and useful suggestions. We also thank S. Chen and Z. Xiao for providing us
with the numerical data used in the appendix. We acknowledge helpful comments and suggestions
by two anonymous referees. This work was carried out under the auspices of the National Nuclear Security Administration of the U.S. Department of Energy at Los Alamos National Laboratory under Contracts W-7405-ENG-36 and DE-AC52-06NA25396.

\section{Appendix }
In this appendix, we show that our results and conclusions about the enstrophy cascade 
are not sensitive to the choice of filtering kernel, $G_\ell(\br)$, and its localization in x-space and k-space.
We also describe in more detail the criterion used in filtering our data in the presence of domain boundaries.
Finally, we show that our results are not corrupted by limited data resolution, present in any experiment including ours,
when considering the cascade across scales larger than the resolution cutoff.

\subsection{Different Filters \label{subsec:DifferentFilters}}

\begin{figure}
  \includegraphics[width=3in]{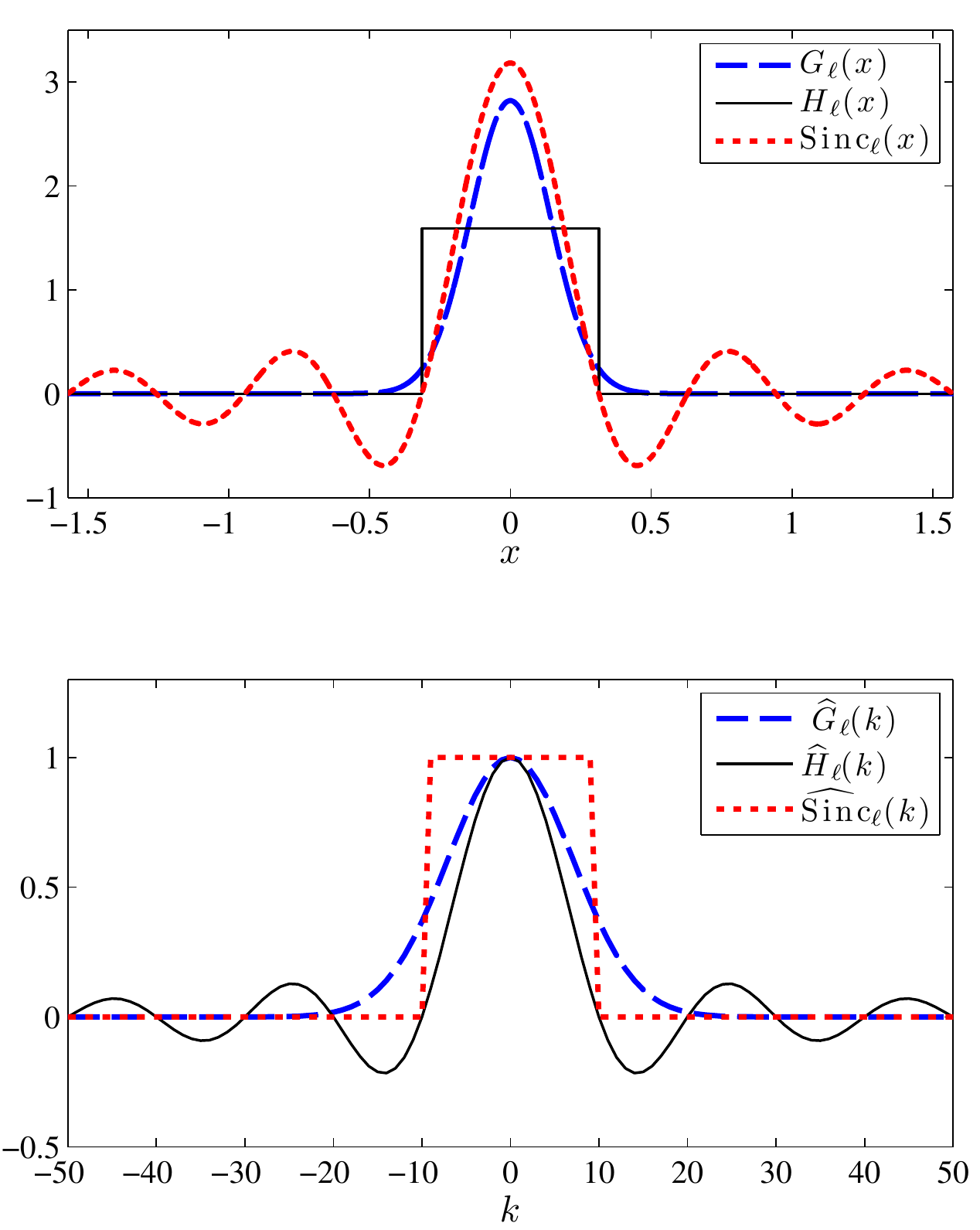}
  \caption{ Comparing the Gaussian, Top-hat, and Sharp-spectral filters 
  in physical (top-panel) and Fourier (bottom-panel) space. See Eqs. (\ref{app_eq:Gauss_x})-(\ref{app_eq:SharpSpectral_k}).
All kernels are normalized in x-space and have width ${\mathcal O}(\ell)$, where $\ell = L/k_\ell$, $k_\ell = 10$, and $L=2\pi$. 
In k-space, all kernels approach $1$ as $k\to0$, retaining low wavenumbers intact while tapering off higher modes.
The more localized a filter is in x-space, the more spread it has to be in k-space and vice versa.
  \label{fig:Filters_xk_space} }
\end{figure}

The interpretation of our results using the coarse-graining approach may depend 
on the functional form of the filter kernel, $G_\ell(\br)$. In our analysis in Section \ref{sec:Results},
we used a Gaussian kernel, $G_\ell(r) = (\pi/\ell^2) \exp{(-\pi^2 r^2 / \ell^2)}$, and its Fourier
transform, $\widehat{G}_\ell(k)$, expressed in Eq. (\ref{eq:GaussianKernel}), which is also a Gaussian. 
Some advantages of a Gaussian kernel are:\\
\begin{enumerate}
\item It is positive, $G_\ell(r)> 0$, in physical space making it easy to interpret the filtering operation (\ref{filtering}) as a local space average in a circle centered at point $\bx$ and with radius $\mathcal{O}(\ell)$.

\item It is relatively compact in physical space, decaying rapidly with $r\to \infty$. This is important in making proper sense of the relevant dynamical quantities, such as the flux $Z_\ell(\bx)$, locally in space. In other words, the Gaussian kernel ensures that there is virtually no contribution to $Z_\ell(\bx)$ from the flow beyond a distance $\mathcal{O}(\ell)$ from $\bx$.

\item Its Fourier transform, $\widehat{G}_\ell(k)$, is also relatively compact with a radius of $\mathcal{O}(k_\ell)$ in Eq. (\ref{eq:GaussianKernel}), decaying rapidly with $k\to\infty$ (see Figs. \ref{fig:SharpGaussFilters} and \ref{fig:Filters_xk_space}). This is important in relating scale $\ell$ in physical space to a wavenumber $k$,  and in comparing our results to more traditional diagnostics that are defined in Fourier space.
\end{enumerate}

The coarse-graining framework is not restricted to a particular choice of a filter kernel. 
Although some traditional turbulence diagnostics have relied on analyzing the flow in Fourier
space, these are only a special case within the coarse-graining approach for a particular choice
of kernel (see Eqs. (\ref{app_eq:SharpSpectral_x}),(\ref{app_eq:SharpSpectral_k}) below).  
The powerful advantage of coarse-graining over traditional Fourier-based diagnostics, however, 
is an ability to probe the flow both in scale and in space.

Owing to the uncertainty principle, the more localized a filter is in x-space, the more spread it has to be
in k-space and vice versa. We demonstrate this well-known fact in Fig. \ref{fig:Filters_xk_space}, where we show 
in 1D three filters that are often used in the LES literature\cite{SagautBook00}: Gaussian, Top-hat, and Sharp-spectral filters. 
The analytical expressions of a normalized Gaussian kernel of width ${\mathcal O}(\ell)$ and its Fourier transform are, respectively:
\begin{eqnarray}
G_\ell(x) &=& \left(\frac{\pi}{\ell^2}\right)^{1/2} e^{-\pi^2 x^2/\ell^2},\lb{app_eq:Gauss_x}\\
\widehat{G}_\ell(k) &=&  e^{-k^2/k_\ell^2}\lb{app_eq:Gauss_k}.
\end{eqnarray}
A normalized Top-hat kernel of width $\ell$ and its Fourier transform are, respectively:
\begin{eqnarray}
H_\ell(x)&=&\begin{cases}
    1/\ell, & \text{if $|x|<\ell/2$}.\\
    0, & \text{otherwise}.\\
  \end{cases}\lb{app_eq:Tophat_x}\\
\widehat{H}_\ell(k) &=& \frac{\sin(\pi k/k_\ell)}{\pi k/k_\ell} \lb{app_eq:Tophat_k}
\end{eqnarray}
The Sharp-spectral filter, which is a top-hat function in k-space, is a normalized $\sinc$ function of width ${\mathcal O}(\ell)$ in x-space:
\begin{eqnarray}
\sinc_\ell(x)&=&\frac{2}{\ell}\frac{\sin(2\pi x/\ell)}{2\pi x/\ell}\lb{app_eq:SharpSpectral_x}\\
\widehat{\sinc}_\ell(k) &=&\begin{cases}
    1, & \text{if $|k|<k_\ell$}.\\
    0, & \text{otherwise}.\\
  \end{cases}\lb{app_eq:SharpSpectral_k}
\end{eqnarray}
Notice that the Top-hat kernel is very well localized in x-space, having a clearly defined width $\ell$.  
Its Fourier transform is very spread in k-space, however, decaying slowly as $\sim k^{-1}$. On the other hand, 
the Sharp-spectral filter is well-localized in k-space but is spread in x-space, decaying like $\sim x^{-1}$. The 
Gaussian kernel offers a reasonable compromise between the Top-hat and Sharp-spectral filters, decaying 
rapidly in both physical and Fourier space.

From an experimental point of view, kernels localized in physical space are better
suited than those localized in Fourier space. This is because
experimental data is invariably windowed to the cross-section of the
measurement apparatus, limiting the extent of the domain to be analyzed.
Kernels such as Eq. (\ref{app_eq:SharpSpectral_x}) of a Sharp-spectral filter 
are poorly localized in x-space and will extend beyond the boundaries of the domain, 
making it hard to interpret the diagnostics. This is not a problem for filters
well-localized (or compact) in x-space. Yet, it is still important to compare 
to theoretical and numerical results which, in turbulence, have traditionally relied
on Fourier analysis. See, for example the standard discussion in Frisch\cite{Frisch_Turbulence},
which relies on a sharp truncation of the Fourier series. 
For such comparisons, kernels such as eq. (\ref{app_eq:SharpSpectral_x})
corresponding to the Sharp-spectral filter are better suited.

\subsubsection{Sensitivity of Results}
We now investigate the sensitivity of our enstrophy cascade results
 to the localization of filters in x-space and k-space. It would be inappropriate
 to use the Sharp-spectral filter in Eqs. (\ref{app_eq:SharpSpectral_x})-(\ref{app_eq:SharpSpectral_k})
 in our sensitivity analysis owing to its poor localization and the finiteness of our domain. This problem
 would vanish for a very large domain and we consider it in more detail when discussing
 the role of boundaries below. Here, we utilize a different set of kernels.

\begin{figure}
 \includegraphics[width=3.2in]{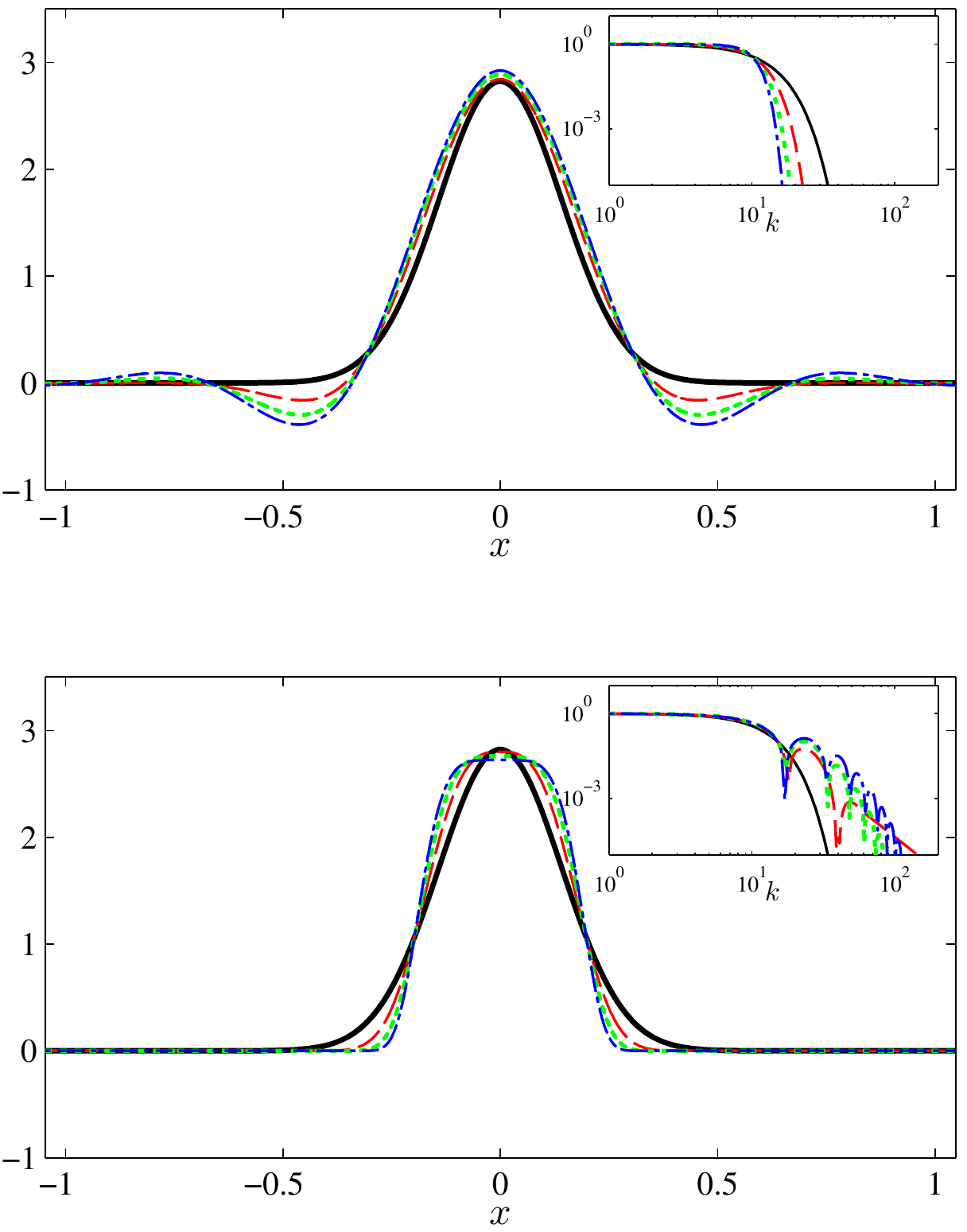}
   \caption{ Plots in x-space of the 
    (upper panel) inverse Fourier transform of $\widehat{G}^{(n)}_\ell(\bk)$ in eq. (\ref{eq:sharpfourier}),
 and (lower panel) of $G^{(n)}_\ell(\bx)$ in eq. (\ref{eq:sharpreal}), where $\ell = L/k_\ell$, $k_\ell = 10$, and $L=2\pi$. 
Insets show square-root of the Fourier spectrum of the respective kernels.
Line style corresponds to the localization-order: 
 n=2 (black solid),  n=3 (red dash), n=4 (green dotted), n=5 (blue dash-dot).
   \label{fig:FilterComp_localize} }
\end{figure}

To make filters more localized in k-space, we use 
\begin{equation}
\widehat{G}^{(n)}_\ell(\bk)=e^{-\left( |\bk|/k_\ell \right)^n},
\label{eq:sharpfourier}
\end{equation}
where $n$ is the localization order, as Fig. \ref{fig:FilterComp_localize} illustrates. 
For $n=2$, expression (\ref{eq:sharpfourier}) reduces to 
the standard Gaussian filter considered earlier.  As $n$ increases, the filter sharpens in k-space
around wavenumber $k_\ell$. In the limit $n\to \infty$, $\widehat{G}^{(n)}_\ell(\bk)$ 
becomes a Sharp-spectral filter expressed in Eq. (\ref{app_eq:SharpSpectral_k}).
To localize filters in x-space, we use
\begin{equation}
G^{(n)}_\ell(\bx)=A~e^{-\left( \pi |\bx|/\ell \right)^n},
\label{eq:sharpreal}
\end{equation}
where $A$ is a normalizing factor. Again, for $n=2$, expression (\ref{eq:sharpreal}) reduces to 
the standard Gaussian filter considered earlier. In the limit $n\to\infty$, $G^{(n)}_\ell(\bx)$ becomes
a Top-hat filter of width  $2\ell/\pi$.  
The two sets of filters (\ref{eq:sharpfourier}) and (\ref{eq:sharpreal}) are shown in x-space in 
Fig. \ref{fig:FilterComp_localize}. 

$$$$
$$$$
$$$$
$$$$
\begin{figure*}[!htbp]
  \includegraphics[width=2.in]{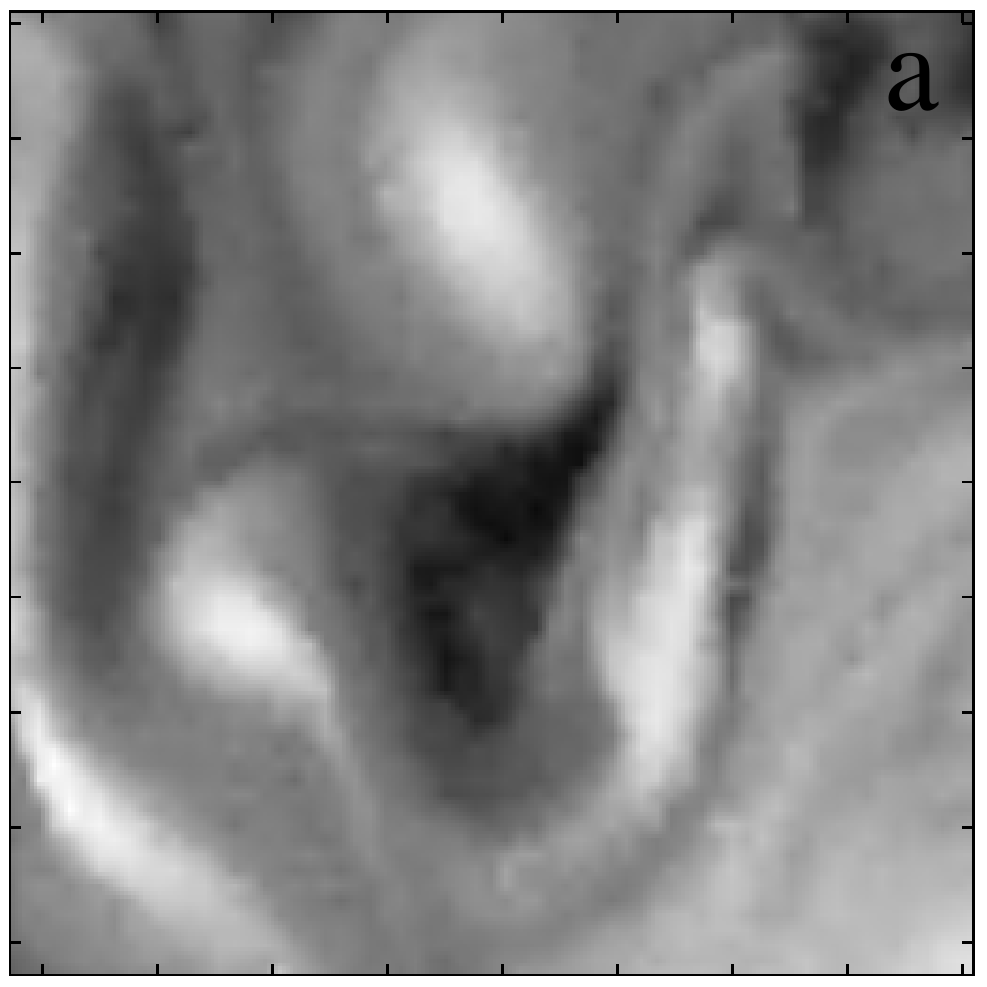}
  \includegraphics[width=2.in]{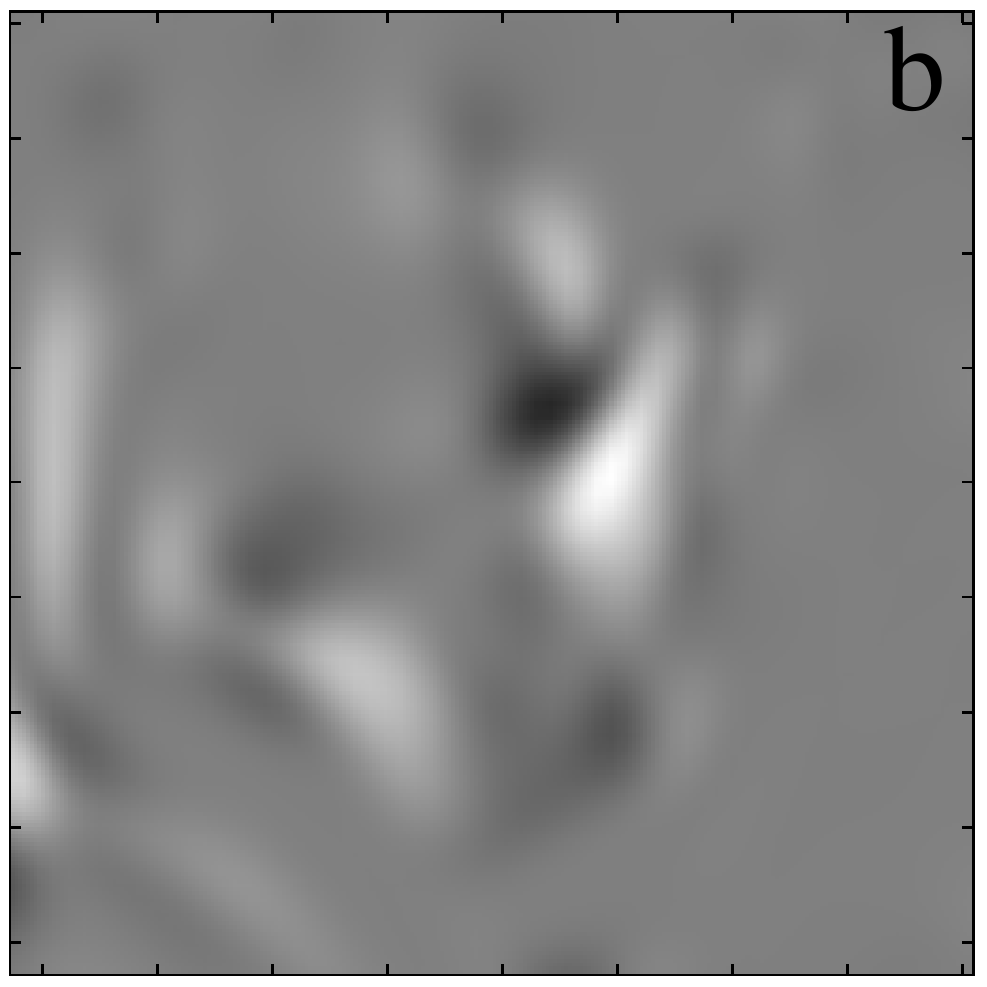}
  \includegraphics[width=2.in]{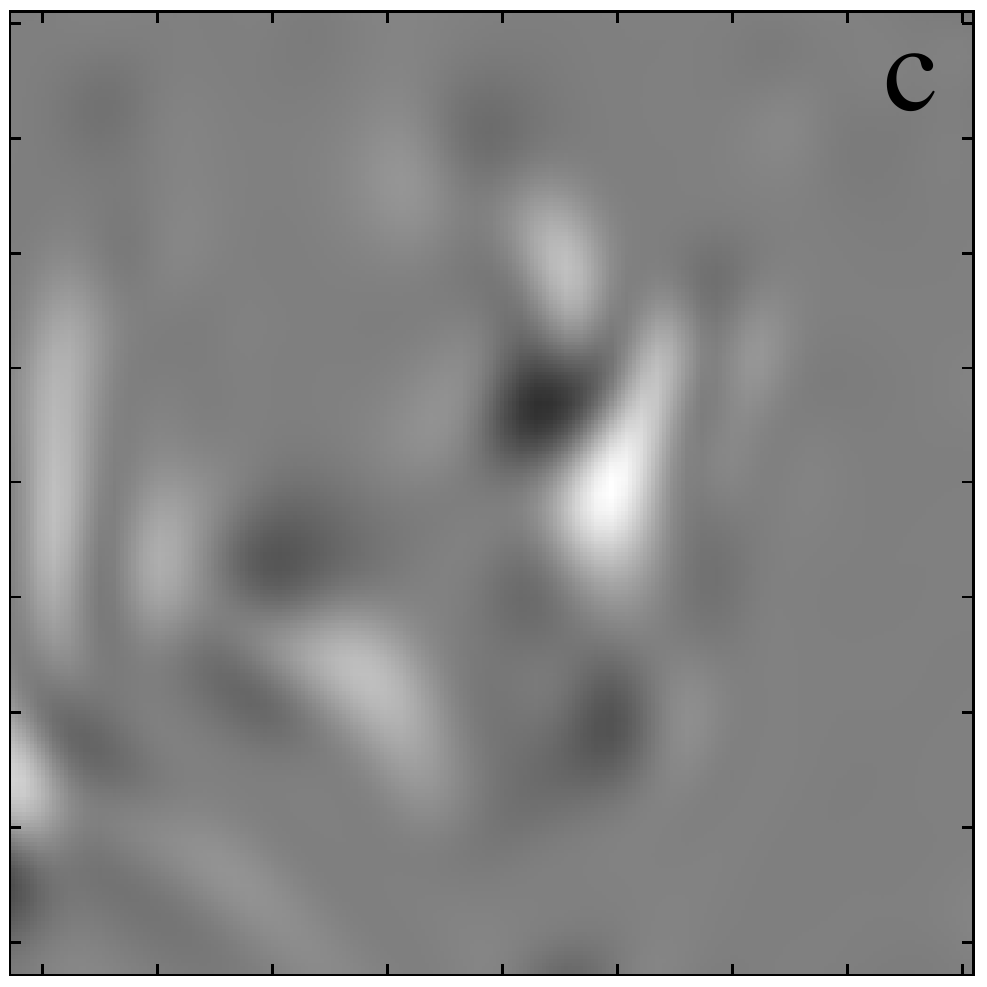}
  \includegraphics[width=2.in]{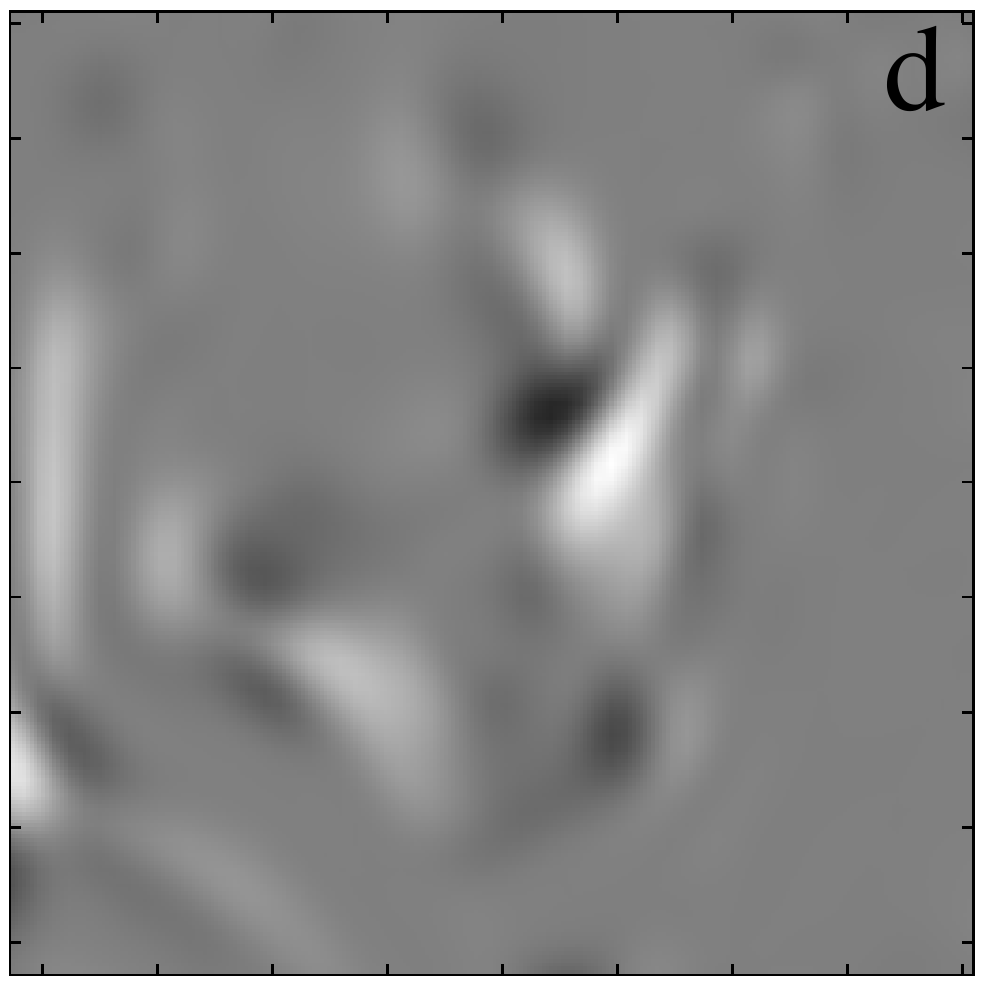}
  \includegraphics[width=2.in]{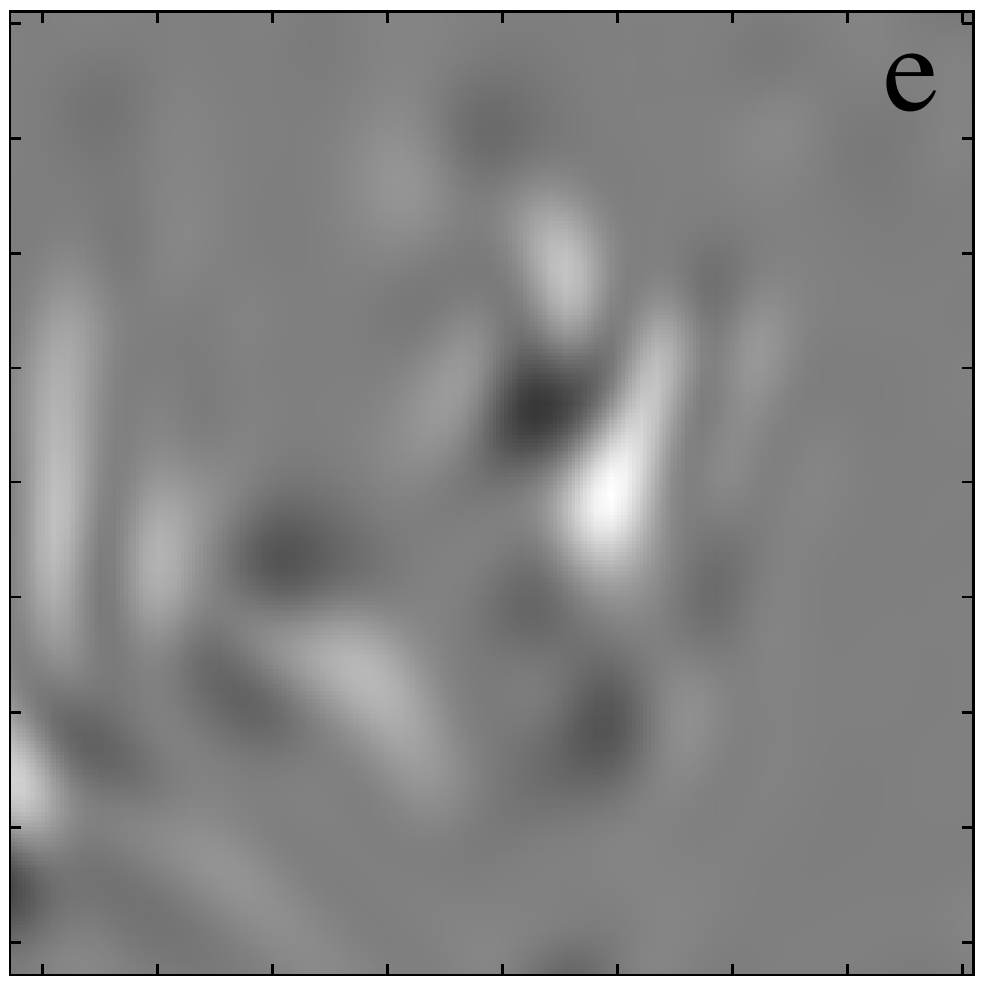}
  \includegraphics[width=2.in]{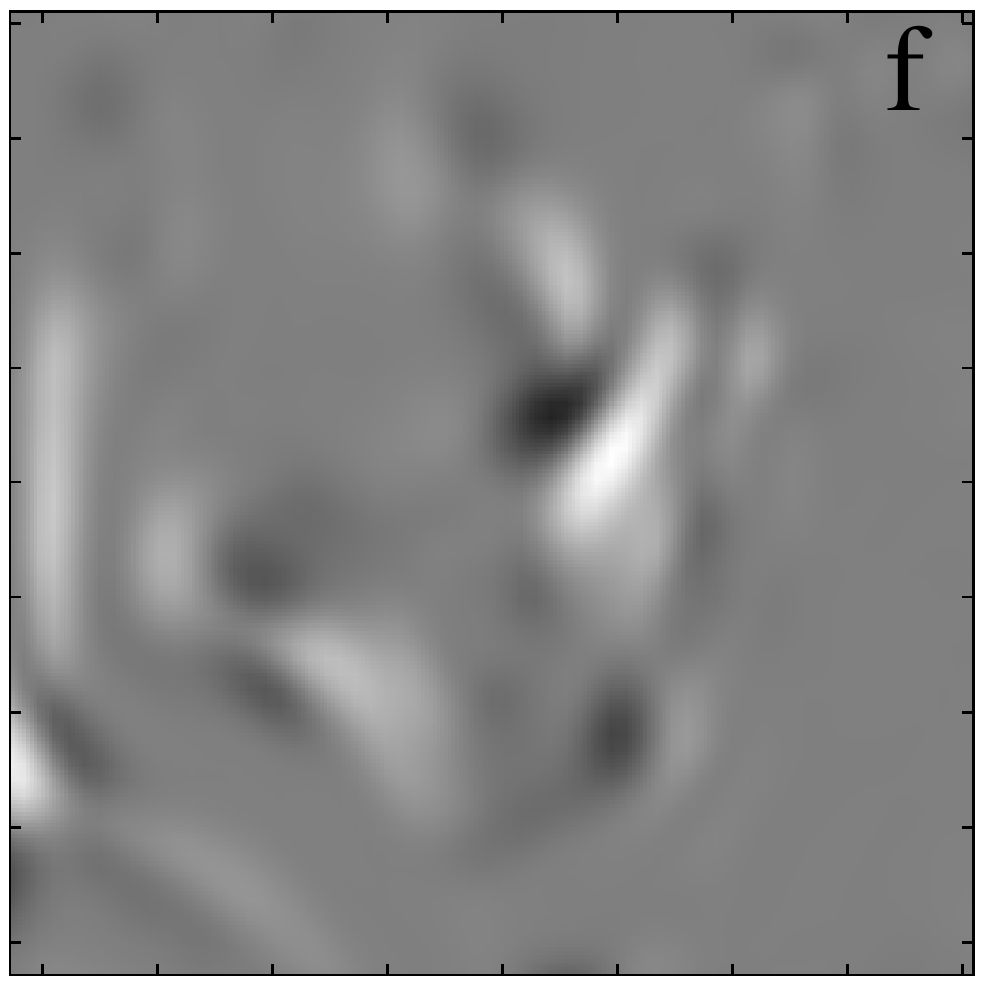}
  \caption{(a) Subsection of a typical vorticity field obtained from
  the soap film. (b)-(f) Calculation of $Z_{\ell}(\bx)$ from vorticity field shown
  in (a) using $\ell=0.2$ cm and various Fourier and real space filters. (b)
  Gaussian filter (grayscale range $\pm 1.95 \times 10^7$s$^{-3}$) (c)
  Fourier filter of order $n=3$ (grayscale range $\pm 2.47 \times
  10^7$s$^{-3}$) (d) Real filter of order $n=3$ (grayscale range $\pm 1.93
  \times 10^7$s$^{-3}$) (e) Fourier filter of order $n=4$ (grayscale range
  $\pm 2.83 \times 10^7$s$^{-3}$) (f) Real filter of order $n=4$ (grayscale
  range $\pm 1.90 \times 10^7$s$^{-3}$).  The hatch marks represent $1$ mm
  increments. \label{fig:FilterComp_Zviz} }
\end{figure*}

\begin{figure}
  \includegraphics[width=3.2in]{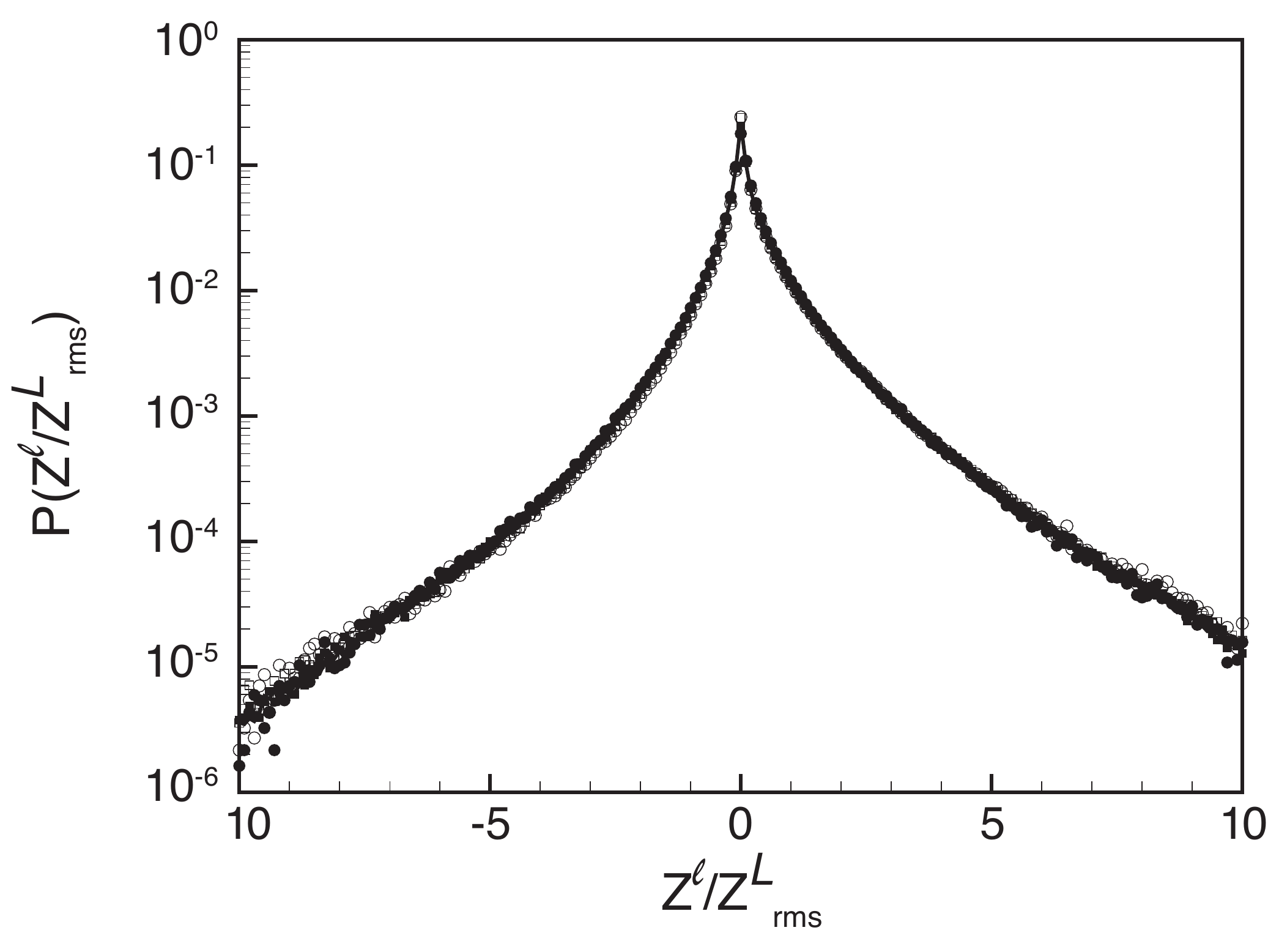}
  \caption{ The probability density function (normalized by RMS) for
  $Z^{\ell}$ obtained using a Gaussian (solid line) two Fourier-space filters
  of order $n=3$(solid squares) and $n=4$ (solid circles) and two real-space
  filters of order $n=3$(open squares) and $n=4$ (open circles). The filter
  length is $\ell=0.2$ mm.
  \label{fig:FilterComp_Zpdf} }
\end{figure}

\begin{figure}
  \includegraphics[width=3.2in]{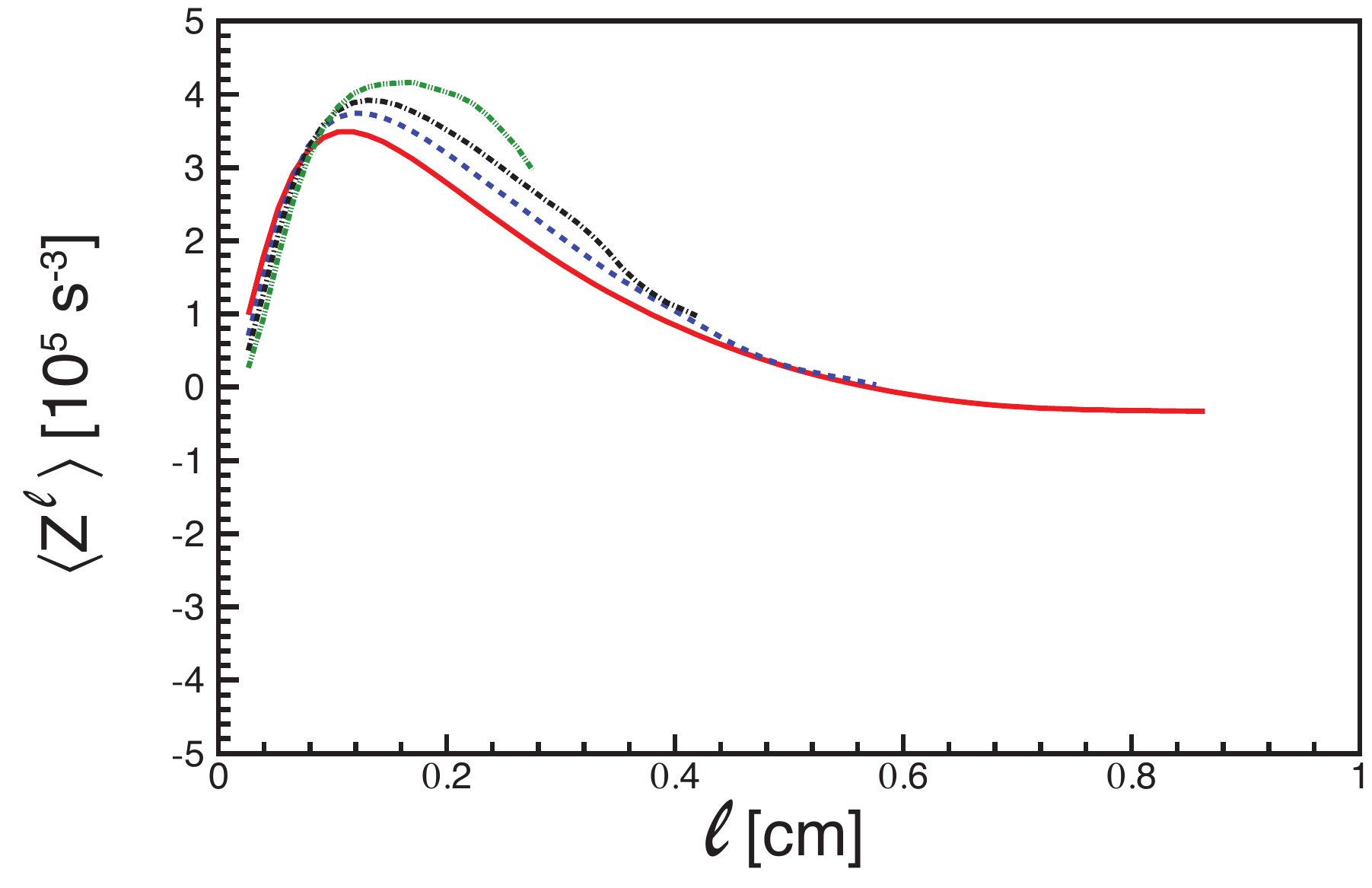}
  \includegraphics[width=3.2in]{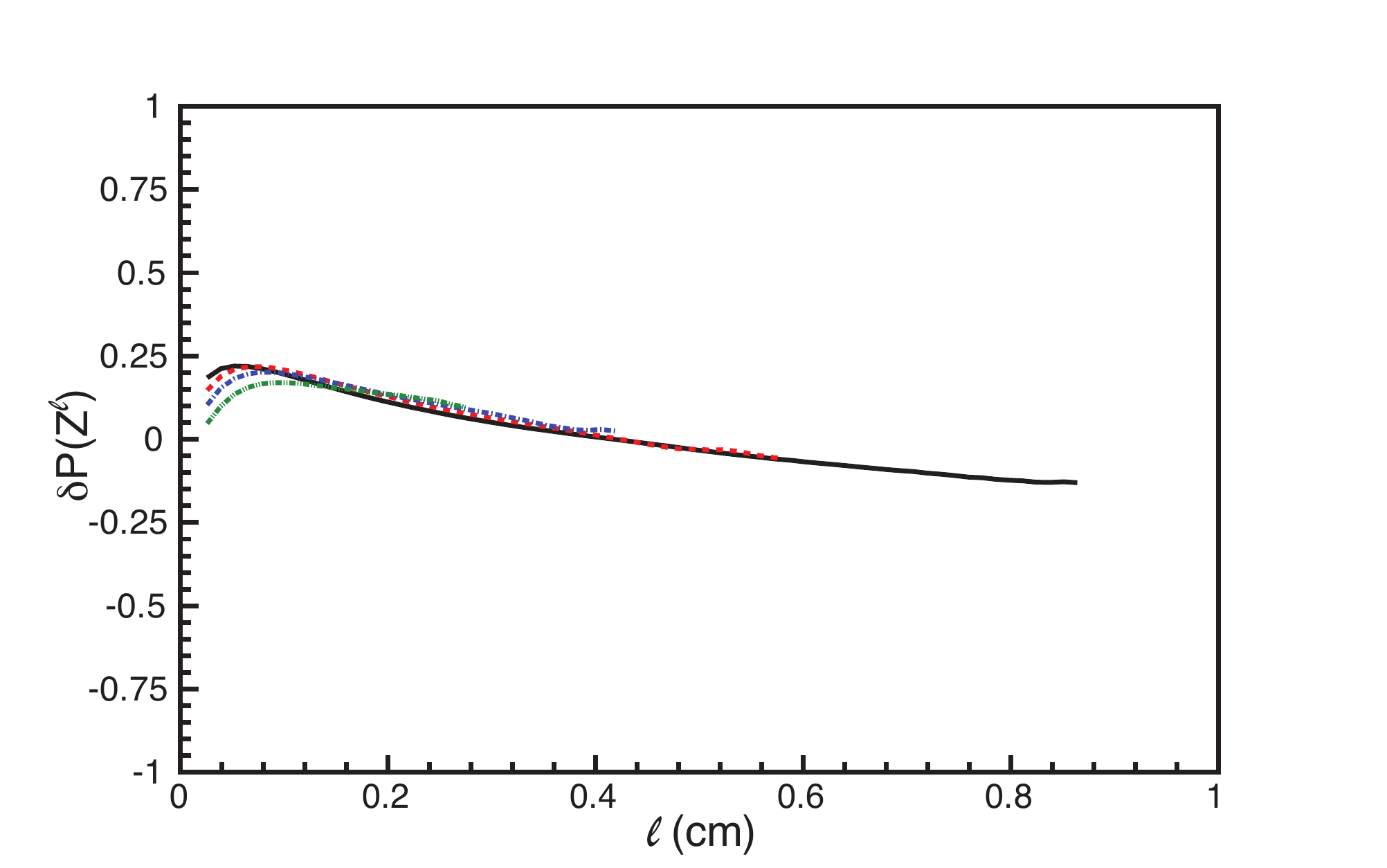}
  \caption{
  (Left Panel) Average enstrophy flux $\langle Z^{\ell} \rangle$
  and (Right Panel) $\delta P(Z^{\ell})$ (see text) for a range of length scales
  calculated using a Gaussian filter
  (solid), and filters of decreasing localization in x-space from eq. (\ref{eq:sharpfourier})
  with $n=3$ (dashed), $n=4$ (dashed-dotted), $n=5$ (dashed-dotted-dotted).
   \label{fig:FilterComp_Zavg} }
\end{figure}

Using the soap-film data, we show in Fig.~\ref{fig:FilterComp_Zviz} visualizations of the enstrophy cascade
$Z_\ell(\bx)$ using filters (\ref{eq:sharpfourier})-(\ref{eq:sharpreal}) for $n=2,3,4$. 
Qualitatively, the fields $Z_\ell(\bx)$ are almost identical. Quantitatively, as indicated by the color bar,
we find a slight increase in the extreme values for filters more localized in k-space (or more spread in x-space).
Overall, however, the qualitative features of $Z_\ell(\bx)$ appear to be insensitive to filter localization.
This is important for correlating the cascade with topological features in a flow.

For a more quantitative comparison, we show the spatial distribution of 
the enstrophy cascade in  Fig.~\ref{fig:FilterComp_Zpdf}. We find that 
all filters considered yield almost identical probability density functions (pdf).
This is consistent with visualizations in Fig. \ref{fig:FilterComp_Zviz}
and reinforces our conclusion that $Z_\ell(\bx)$ is fairly insensitive to 
filter choice.

To further assess the effect filter choice has on the cascade field,
we show in Fig. \ref{fig:FilterComp_Zavg} the spatial average,
$\langle Z_\ell \rangle$ and fractional sign-probability, 
$\delta P(Z^{\ell}) \equiv P(Z^{\ell}>0) - P(Z^{\ell} < 0)$, as a function of scale $\ell$.
Although plots of $\delta P(Z^{\ell})$ using different filters are fairly similar, we find a
noticeable difference in plots of $\langle Z_\ell \rangle$. As we mentioned in the beginning of
section \ref{sec:Results}, this difference is to be expected since the more spread a filter is in k-space, 
the more averaging it entails over scales (see Fig. \ref{fig:SharpGaussFilters}), thus 
yielding a smoother plot of $\langle Z_\ell \rangle$ as a function of $\ell$, typically with 
smaller values. These differences between filters will vanish if we keep increasing the range of scales between
enstrophy injection and dissipation since $\langle Z_\ell \rangle$ becomes constant as a function of $\ell$ and, hence, 
insensitive to any averaging in scale.

\subsection{Boundaries \label{subsec:Boundaries}}

We have made sure that all results in this paper are not affected by the 
domain boundary due to the viewing window, as we shall now explain.
The filtering operation, as defined in Eq. (\ref{filtering}), requires information about the
flow within a distance ${\mathcal O}(\ell)$ around any location $\bx$ to be analyzed. The 
precise distance needed depends on the physical-space localization of the filter, as discussed 
in appendix \ref{subsec:DifferentFilters}. The presence of a boundary within this distance away from
$\bx$ produces errors in the derivation of the coarse-grained equations (\ref{largeVelocityEq})-(\ref{largeVorticityEq}).
These so-called ``commutation errors'' are an active research subject in LES modeling and 
arise because the filtering operation (\ref{filtering}) and spatial derivatives no
longer commute, $\OL{\grad f} \ne \grad \OL{f}$ (see Chapter 2 in Ref\cite{SagautBook00} for more details).

\begin{figure}
   \includegraphics[width=3.2in]{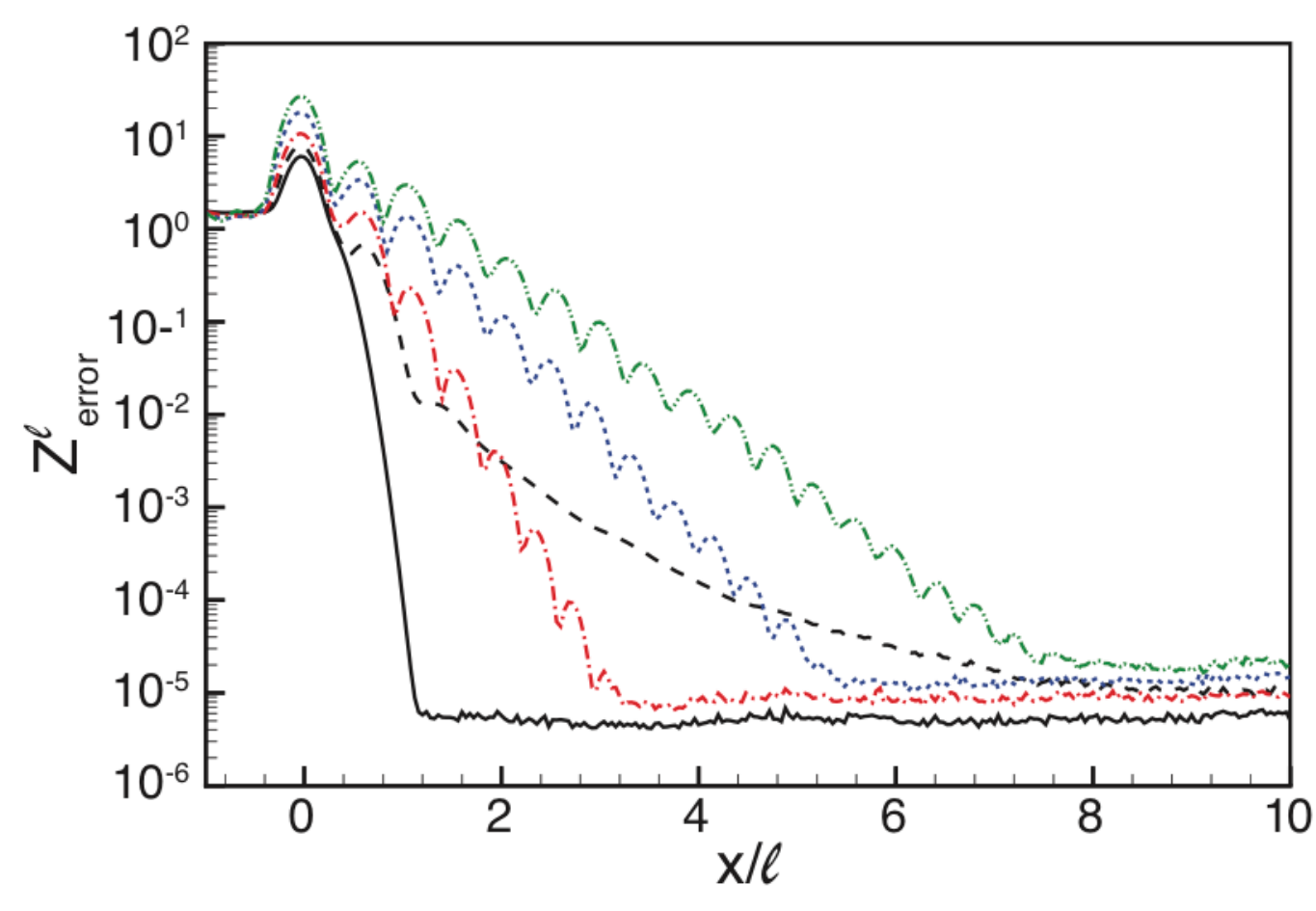}
   \caption{
Error in calculating the cascade, defined in eq. (\ref{eq:BoundaryError}), as a result of filtering data a distance $x/\ell$ from the boundary. Plots are for filters (\ref{eq:sharpfourier}) with decreasing spatial localization of order 
$n=2$, which is a Gaussian (solid line), $n = 3$ (dashed), $n = 4$ (dash-dot), $n = 6$ (dotted), $n = 8$ (dash-dot-dot)
 \label{fig:FilterComp_Zboundary} }
\end{figure}

To gauge the effect boundaries could have on calculating the cascade field if 
they are not avoided as we did in this paper, we used numerical data\footnote{Numerical data was kindly provided to us by S. Chen and Z. Xiao (private communication). See Ref.\cite{Chenetal03} for simulation details.\label{NumericalFootnote}}
a flow field in a doubly periodic domain on a $512\times512$ grid. We then
computed the enstrophy flux field, $Z^{\ell}_0(\bx)$, using
filters (\ref{eq:sharpfourier}) for different values of $\ell$.
To mimic the presence of a viewing window, we set the flow to zero in half of 
the domain and recomputed the flux field, $Z^{\ell}_b(\bx)$. The resultant root-mean-square
of the difference,
\begin{equation}
Z^{\ell}_\text{error}(x) = \frac{\langle (Z^{\ell}_b - Z^{\ell}_0)^2
\rangle_y^{1/2}}{ (Z^{\ell}_0)_\text{rms}},
\lb{eq:BoundaryError}\end{equation}
is shown in Fig.~\ref{fig:FilterComp_Zboundary} as a function of
distance from the introduced boundary. The spatial averaging, $\langle\dots\rangle_y$, in Eq. (\ref{eq:BoundaryError})
is along the y-direction, parallel to the boundary. 

Fig. \ref{fig:FilterComp_Zboundary} shows the error arising from different filters (\ref{eq:sharpfourier}) of 
decreasing localization in x-space. This corresponds to an increasing $n$ in Eq. (\ref{eq:sharpfourier})
(see also upper panel in Fig. (\ref{fig:FilterComp_localize})).
As expected, we find that the error increases for filters less localized in x-space. The Gaussian filter with $n=2$, 
which is the most localized, has the smallest error. 

For the results presented in this paper, we have considered a maximum error rate of $10^{-4}$ as acceptable.
Based on Fig. \ref{fig:FilterComp_Zboundary}, this stipulates that we restrict our analysis to flow locations at least 
a distance $\approx \ell (n-1)$ from the domain boundary, where $n$ is the localization order in Eq. (\ref{eq:sharpfourier}). 
Filters (\ref{eq:sharpreal}) that are more localized than a Gaussian do not lead to such a severe loss of data.

\subsection{Finite Measurement Resolution \label{subsec:FiniteResolution}}

\begin{figure}
  \includegraphics[width=3.2in]{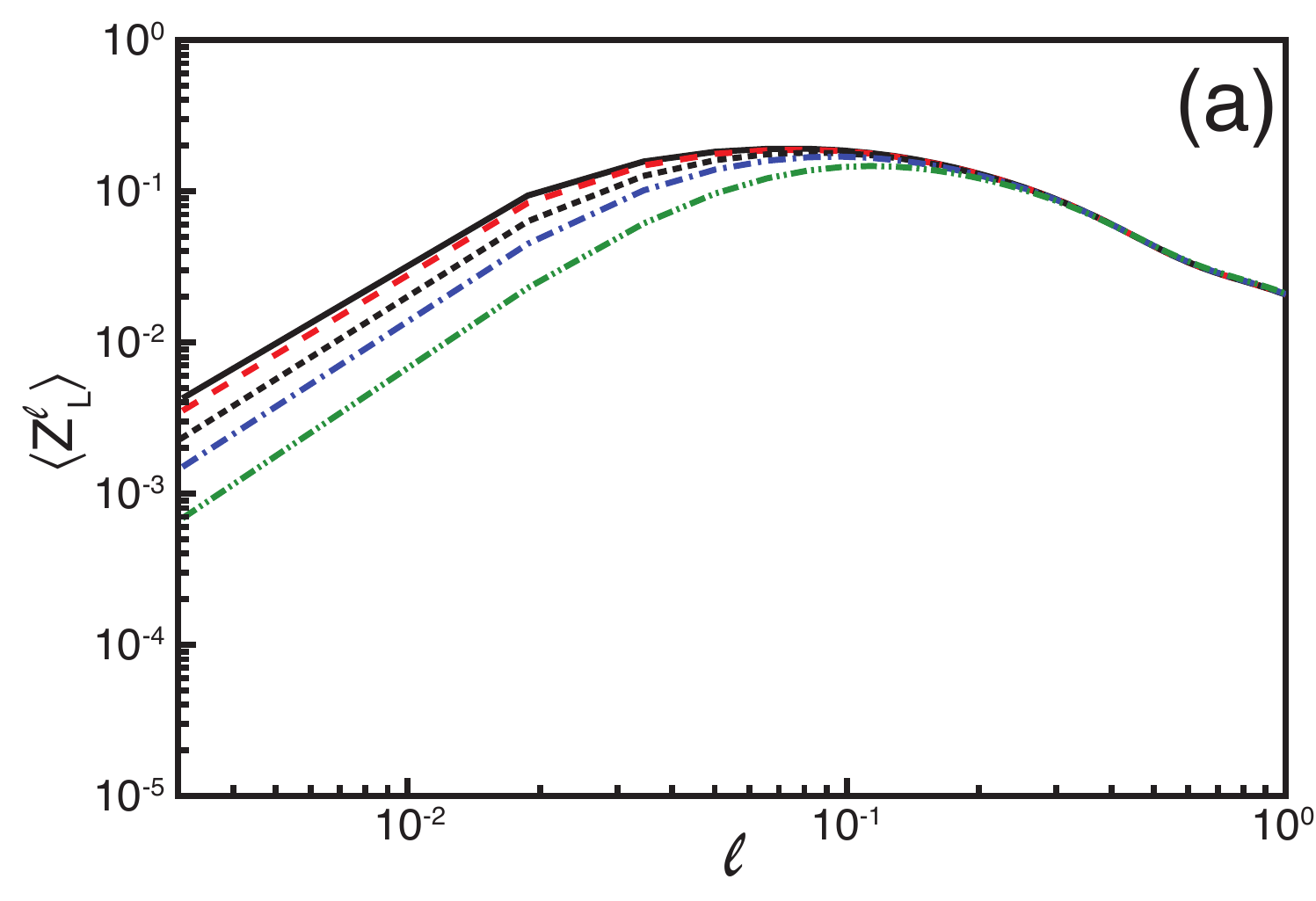}
  \includegraphics[width=3.2in]{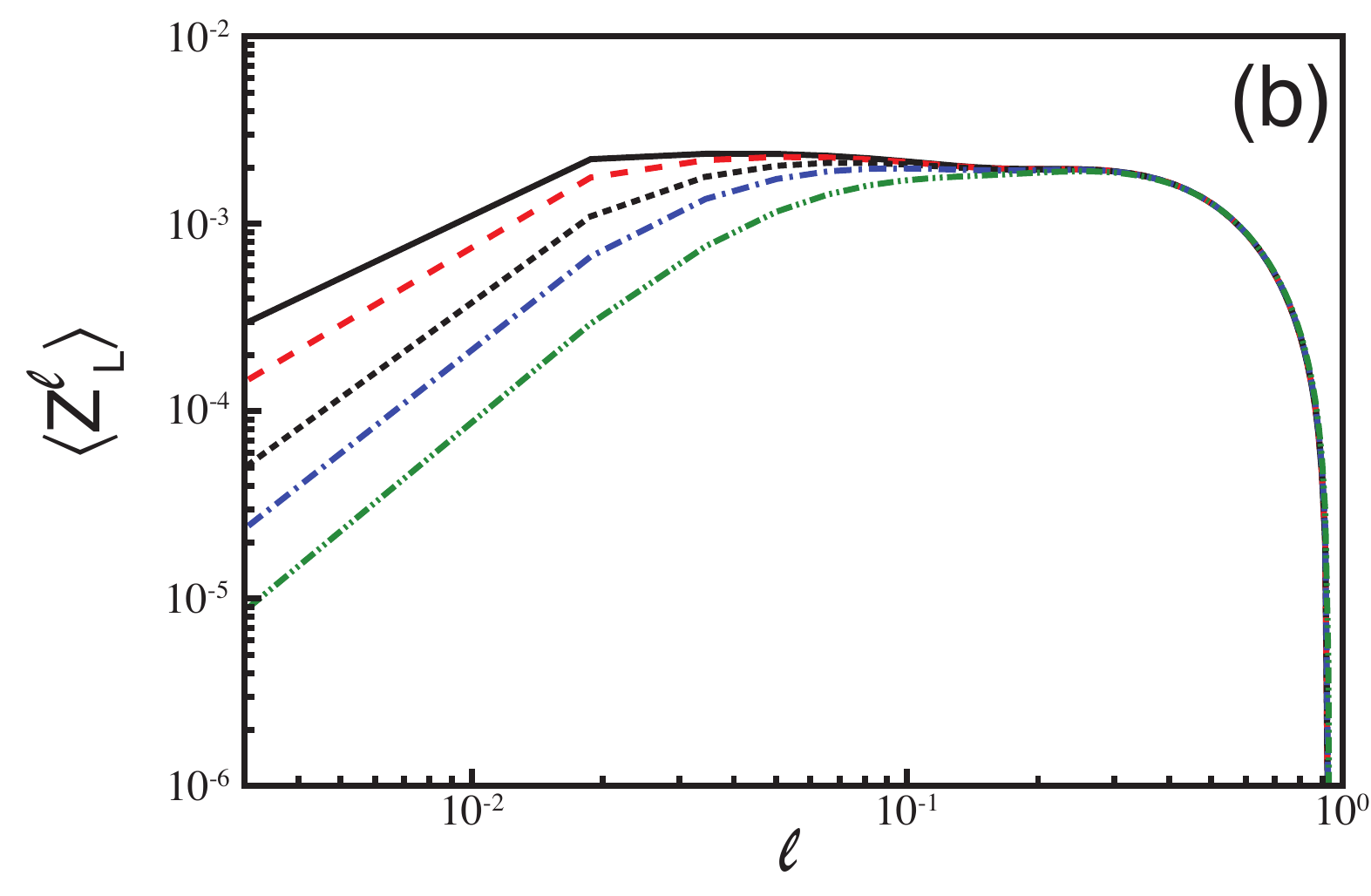}
  \caption{
Mean enstrophy flux from numerical data
using Laplacian viscosity (left panel) and hyper-viscosity (right panel).
Before computing the flux, the velocity field is pre-filtered at scale $L = 0.01$ (dashed), 0.02 (dotted), 0.03 (dash-dot), 0.05 (dash-dot-dot). The total enstrophy flux  without pre-filtering (solid line) is also shown for comparison. 
The filter (\ref{eq:sharpreal}) used is highly localized in x-space, with $n=20$, similar to a Top-hat filter.
  \label{fig:ResolutionAnalysis} }
\end{figure}

We now explore the effect of finite measurement resolution on determining the value of 
mean enstrophy cascade. This issue is common to both the coarse-graining and traditional 
Fourier-based techniques. To this end, we use doubly periodic 2D simulations 
which, unlike soap-film data presented in the paper, have a well-controlled range of scales that is well-resolved
by the numerical grid and free from any boundary effects. We utilize two sets of simulations\cite{Chenetal03}, 
one with a normal Laplacian viscous term, $\nu_1 \nabla^2 \bu$, and another with 
hyper-viscosity, $\nu_8 \nabla^{16} \bu$.

To mimic the impact of limited measurement resolution, we pre-filter the velocity field using kernel
$G^{(n)}_\ell$, defined in eq. (\ref{eq:sharpreal}), that is highly localized in x-space, with $n=20$, 
similar to a Top-hat filter. As expected, Fig. \ref{fig:ResolutionAnalysis} shows that pre-filtering
at a scale $L$ comparable to the grid cell-size $\Delta x$ has minimal effect on the 
computed enstrophy flux values, $\langle Z_\ell\rangle$, especially when considering 
the cascade across $\ell \gg L$. As the pre-filter length, $L$, is increased 
the computed enstrophy flux decreases in value over a larger range of scales.  
The effects of pre-filtering, however, are felt most acutely at scales $\ell \lesssim L$. There is little change 
in the flux $\langle Z_\ell\rangle$ values at scales $\ell \gg L$. 

The problem of limited measurement resolution is intimately related to scale-locality of the enstrophy 
cascade in 2D turbulence. Our results are consistent with theoretical predictions due to Kraichnan\cite{Kraichnan67} 
and Eyink \cite{Eyink05}, who argued that the nonlocal nature of the enstrophy cascade across
a scale $\ell$ is from the influence of much larger scales $\Delta\gg \ell$, where $\Delta$ is 
typically the scale at which the energy spectrum peaks. The authors specifically argued that the 
influence of much smaller scales, $\delta \ll \ell$, have a negligible contribution to the cascade across
$\ell$. This is consistent with our observations in Fig. \ref{fig:ResolutionAnalysis} that there is a negligible effect of 
pre-filtering at scale $L$ on the value of the cascade across scales $\ell \gg L$. 

\bibliography{2D_Turbulence}

\end{document}